\pdfoutput=1
\documentclass[11pt]{article}

\usepackage{acl}

\usepackage{times}
\usepackage{latexsym}

\usepackage[T1]{fontenc}
\usepackage[utf8]{inputenc}

\usepackage{microtype}

\usepackage{inconsolata}

\usepackage{amsmath} 
\usepackage{booktabs}
\usepackage{float}
\usepackage{graphicx}
\usepackage{makecell}
\usepackage{multirow}
\usepackage{paralist}
\usepackage{subfig}
\usepackage{url}
\usepackage{xspace}
\usepackage[linesnumbered,ruled,vlined]{algorithm2e}
\usepackage{listings}
\newcommand{\methodname}{\textsc{DraCo}\xspace}
\newcommand{\crosscodeeval}{CrossCodeEval\xspace}
\newcommand{\recceval}{ReccEval\xspace}

\title{Dataflow-Guided Retrieval Augmentation for Repository-Level\\ Code Completion}

\author{
    Wei Cheng$^\dagger$ \quad 
    Yuhan Wu$^\dagger$ \quad 
    Wei Hu$^{\dagger,\,\ddagger,\,}$\thanks{\,\, Corresponding author} \\
    $^\dagger$ State Key Laboratory for Novel Software Technology, Nanjing University, China \\
    $^\ddagger$ National Institute of Healthcare Data Science, Nanjing University, China \\
    \texttt{wchengcs.nju@gmail.com, yhwu.nju@gmail.com, whu@nju.edu.cn} 
}

\begin{document}
\maketitle

\begin{abstract}
Recent years have witnessed the deployment of code language models (LMs) in various code intelligence tasks such as code completion. Yet, it is challenging for pre-trained LMs to generate correct completions in private repositories. Previous studies retrieve cross-file context based on import relations or text similarity, which is insufficiently relevant to completion targets. In this paper, we propose a dataflow-guided retrieval augmentation approach, called \methodname, for repository-level code completion. \methodname parses a private repository into code entities and establishes their relations through an extended dataflow analysis, forming a repo-specific context graph. Whenever triggering code completion, \methodname precisely retrieves relevant background knowledge from the repo-specific context graph and generates well-formed prompts to query code LMs. Furthermore, we construct a large Python dataset, \recceval, with more diverse completion targets. Our experiments demonstrate the superior accuracy and applicable efficiency of \methodname, improving code exact match by 3.43\% and identifier F1-score by 3.27\% on average compared to the state-of-the-art approach.
\end{abstract}

%====================%
\section{Introduction}
Pre-trained language models (LMs) of code \cite{Chen2021Codex,Nijkamp2023CodeGen2,Nijkamp2023CodeGen,Allal2023Santacoder,Li2023Starcoder} have shown remarkable performance in improving programming productivity \cite{Kazemitabaar2023Studying, Dakhel2023GitHub}.
Instead of using a single code file, well-designed programs emphasize separating complicated functionality into independent modules \cite{Barnett1968Modular}.
While facilitating collaborative development and software maintenance, it introduces the real-world problem of \emph{repository-level code completion}: given an unfinished code file in a private repository, complete the following pieces of code at the cursor position.

Despite pre-training on large-scale corpora, code LMs are still blind to unique naming conventions and programming styles in private repositories \cite{Pei2023Better, Liu2023RepoBench, Ding2023CrossCodeEval}.
Previous works finetune LMs to leverage cross-file context \cite{Ding2022CoCoMIC, Shrivastava2023RepoFusion, Shrivastava2023Repository}, which requires additional training data and is difficult to work with larger LMs.
Recently, retrieval-augmented generation (RAG) is widely used to aid pre-trained LMs with external knowledge and maintain their parameters intact \cite{Lewis2020Retrieval, mallen-etal-2023-trust, trivedi-etal-2023-interleaving}.
For repository-level code completion, the retrieval database is the current private repository.
The state-of-the-art approach, RepoCoder \cite{Zhang2023RepoCoder}, iteratively incorporates a text similarity-based retriever and a code LM.

\begin{figure}
    \centering
    \includegraphics[width=\linewidth]{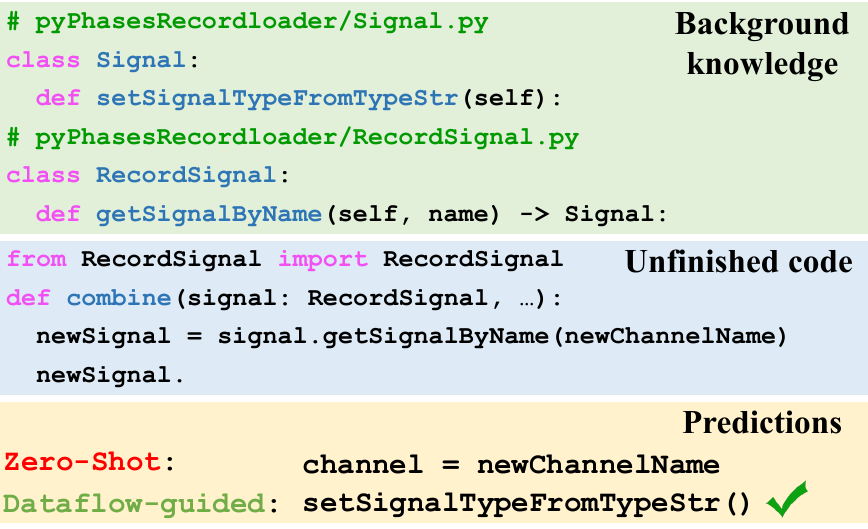}
    \caption{A real-world example of repository-level code completion. 
    The code LM CodeGen25-7B-mono fails to complete the last code line correctly when entering only the unfinished code (Zero-Shot).
    The model needs background knowledge relevant to \texttt{newSignal}, and the retrieval of this knowledge can be guided by dataflow.
    }
    \label{fig:example}
\end{figure}

As shown in Figure~\ref{fig:example}, the CodeGen25 Python model \cite{Nijkamp2023CodeGen2} with 7 billion parameters assigns a value to the attribute \texttt{channel} of the object \texttt{newSignal}, which seems rational in the unfinished code but is actually outside the list of valid attributes.
Due to the lack of similar code snippets in the repository, the text similarity-based approach \cite{Zhang2023RepoCoder} also fails to complete the correct code line.
From a programmer’s perspective, one would explore the data origin of the variable \texttt{newSignal} in the last line.
It comes from the call \texttt{signal.getSignalByName}, where the variable type of \texttt{signal} is \texttt{RecordSignal} imported from the module \texttt{RecordSignal}.
After providing relevant background knowledge in the private repository, the model would know that the variable type of \texttt{newSignal} is the class \texttt{Signal} and thus call the correct function.

Inspired by this programming behavior in private repositories, we propose \methodname, a novel dataflow-guided retrieval augmentation approach for repository-level code completion, which steers code LMs with relevant background knowledge rather than similar code snippets.
Dataflow analysis is a static program analysis reacting to data dependency relations between variables in a program.
In this work, we extend traditional dataflow analysis by setting type-sensitive dependency relations.
We employ the standard RAG framework~\cite{Lewis2020Retrieval}: 
(\romannumeral1) \emph{Indexing}, which parses a private repository into code entities and establishes their relations through dataflow analysis, forming a repo-specific context graph for retrieval.
(\romannumeral2) \emph{Retrieval}, which uses dataflow analysis to obtain fine-grained import information in the unfinished code and retrieves relevant code entities from the pre-built context graph.
(\romannumeral3) \emph{Generation}, which organizes the relevant background knowledge as natural code and concatenates it with the unfinished code to generate well-formed prompts for querying code LMs.

In addition to the existing dataset \crosscodeeval \cite{Ding2023CrossCodeEval} for repository-level code completion, we construct a new dataset, \recceval, with diverse completion targets collected from Python Package Index (PyPI).\footnote{\url{https://pypi.org/}}
We conduct experiments with popular LMs including adapted code LMs \cite{Baptiste2023codellama}, specialized code LMs \cite{Nijkamp2023CodeGen2, Nijkamp2023CodeGen, Allal2023Santacoder, Li2023Starcoder}, and GPT models \cite{Ouyang2022Training, gpt4}.
Our experiments demonstrate that \methodname achieves generally superior accuracy across all settings.
Furthermore, \methodname is plug-and-play for various code LMs and efficient to real-time code completion.

Our main contributions are outlined as follows:
\begin{compactitem}
    \item We design an extended dataflow analysis by setting type-sensitive data dependency relations, which supports more precise retrieval.
    
    \item We propose \methodname, a dataflow-guided retrieval augmentation approach for repository-level code completion.
    \methodname builds a repo-specific context graph for retrieval and generates well-formed prompts with relevant background knowledge in real-time completion.
    
    \item We construct a Python dataset \recceval with diverse completion targets.
    The experimental results show that \methodname improves code exact match by 3.43\%, identifier F1-score by 3.27\%, and prompt generation time by 100$\times$ on average compared to the second-best approach RepoCoder \cite{Zhang2023RepoCoder}.
    Our source code and data are available at \url{https://github.com/nju-websoft/DraCo}.
\end{compactitem}

%====================%
\section{Related Work}

\paragraph{Code completion.} 
Early studies adopt statistical LMs \cite{Raychev2014Code, Proksch2015Intelligent, Raychev2016Probabilistic, He2021PyART} and neural models \cite{Li2018Code, Svyatkovskiy2019Pythia, Kim2021Code, Izadi2022CodeFill, Tufano2023Automating} for code completion.
After pre-training on large-scale code corpora, code LMs are familiar with frequent code patterns and achieve superior performance \cite{Chen2021Codex, Lu2021CodeXGLUE, wang-etal-2021-codet5, Le2022CodeRL, Allal2023Santacoder, Li2023Starcoder, Nijkamp2023CodeGen2, Nijkamp2023CodeGen, Shen2023PanGu, Zheng2023CodeGeeX}.
Unlike single-file code completion, repository-level code completion draws much attention to practical development.
\citet{Ding2022CoCoMIC} learn in-file and cross-file context jointly on top of pre-trained LMs.
\citet{lu-etal-2022-reacc} present ReACC to train a code-to-code search retriever and a code completion generator with an external source code database.
\citet{Shrivastava2023Repository} generate example-specific prompts using a prompt proposal classifier and further propose RepoFusion \cite{Shrivastava2023RepoFusion} to incorporate relevant repository context by training code LMs.
RepoCoder \cite{Zhang2023RepoCoder} is an iterative retrieval-generation framework to approximate the intended completion target.
Despite their good performance, these methods are limited by the high overhead of extra training or iterative generation.

\begin{figure*}
    \centering
    \includegraphics[width=\textwidth]{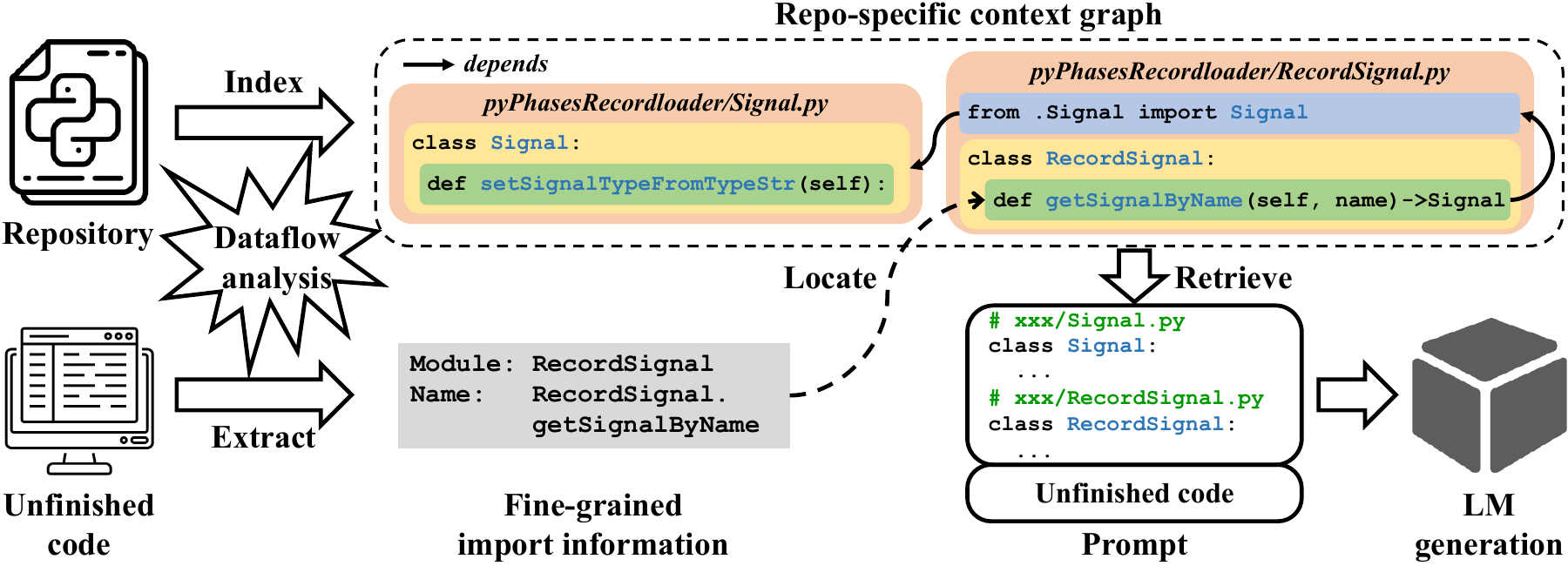}
    \caption{Overview of our approach, where dataflow analysis is crucial for both indexing and retrieval. 
    The details of the unfinished code have been shown in Figure~\ref{fig:example}.
    The rectangular boxes visualize \textit{contains} relations between the code entities in the repo-specific context graph, and the solid arrows indicate \textit{depends} relations. 
    }
    \label{fig:framework}
\end{figure*}

\paragraph{Retrieval-augmented generation.} 
For the scenarios where required knowledge is missing or outdated in pre-trained LMs, RAG has achieved state-of-the-art performance in many NLP tasks \cite{Cai2022Recent, Feng2023Trends, mallen-etal-2023-trust}.
Usually, RAG integrates the retrieved knowledge with frozen pre-trained LMs \cite{Ram2023In, Levine2022Standing, Shi2023REPLUG}.
There exist different types of retrievals including term-based sparse retriever \cite{Robertson2009Probabilistic, trivedi-etal-2023-interleaving}, embedding-based dense retriever \cite{karpukhin-etal-2020-dense, Lewis2020Retrieval}, commercial search engines \cite{Nakano2021WebGPT, Liu2023WebGLM}, and LMs themself \cite{Yu2023Generate, Sun2023Recitation}.
RAG is also broadly applied to code intelligence tasks such as code summarization \cite{Liu2021Retrieval, Zhang2020Retrieval, Zhou2023Towards} and code generation \cite{Hashimoto2018Retrieve, parvez-etal-2021-retrieval-augmented, Li2023Towards}.
In this work, we leverage dataflow analysis to guide retrieval, which mines more precise data dependency information for repository-level code completion.

%====================%
\section{Methodology}

As shown in Figure~\ref{fig:framework}, \methodname employs the standard RAG framework \cite{Lewis2020Retrieval} including indexing (\S\ref{sec:index}), retrieval (\S\ref{sec:retrieval}), and generation (\S\ref{sec:generation}).
Because both indexing and retrieval in \methodname rely on the extended dataflow analysis, we first introduce it in \S\ref{sec:dfa}.
In this work, we focus on Python and the code completion of currently edited line, which simulates real-world scenarios where users are programming in integrated development environments (IDEs) and only the context before the cursor is visible.

\subsection{Dataflow Analysis}
\label{sec:dfa}

Dataflow analysis is a static program analysis that reacts to the data dependency relations between variables in a program, producing a dataflow graph (DFG), in which nodes represent the variables and edges indicate where the variables come from and where they go.
It provides code semantic information that is not affected by personal naming conventions and programming styles.

We assume that the background knowledge relevant to variable types is crucial for code completion.
Take the statement \texttt{v = f(p)} as an example, the parameter \texttt{p} has far less influence on the variable \texttt{v} than the call \texttt{f} does.
Therefore, we extend traditional dataflow analysis by setting dependency relation types.
We focus on five \emph{type-sensitive relations}, which indicate what the variable type is or where it derives from:
\begin{compactitem}
    \item \textit{\underline{A}ssigns} relation is a one-to-one correspondence in an assignment statement, which controls variable creation and mutation.
    
    \item \textit{\underline{As}} relation is from \textit{with} or \textit{except} statements and similar with the \textit{assigns} relation.

    \item \textit{\underline{R}efers} relation represents a reference to an existing variable or its attribute.

    \item \textit{\underline{T}ypeof} relation is from the explicit type hints \cite{pep484} written by programmers, indicating the data type of the (return) value of a variable or function.

    \item \textit{\underline{I}nherits} relation is an implicit data dependency relation since a subclass inherits all the class members of its base classes.
\end{compactitem}

Our DFG is a heterogeneous directed acyclic graph $G = \{(h, r, t) \mid h,t \in E, r \in R\}$, where $E$ denotes the entity set, $R$ denotes the type-sensitive relation set, and a triplet $(h, r, t)$ represents the head entity $h$ pointing to the tail entity $t$ with the relation $r$.
The details of our DFG construction are shown in Appendix~\ref{appendix:dfg}.

\subsection{Repo-specific Context Graph}
\label{sec:index}

Offline preprocessing is often used in RAG to index a retrieval database.
Instead of treating source code as text \cite{lu-etal-2022-reacc, Zhang2023RepoCoder}, we parse a private repository into code entities and establish their relations through our dataflow analysis, forming a repo-specific context graph.

For each code file in a repository, we traverse its abstract syntax tree (AST) to collect code entities including modules, classes, functions, and variables.
A module entity stores its file path and docstring as properties.
A class entity stores its name, signature, docstring, and starting line number.
A function entity stores its name, signature, docstring, body, and starting line number.
A variable entity stores its name, statement, and starting line number.
There are natural \textit{contains} relations between these entities, e.g., the class \texttt{RecordSignal} \textit{contains} its member function \texttt{getSignalByName}.
Based on the type-sensitive relations in DFG, we establish \textit{depends} relations between the entity pairs in individual modules, e.g., \texttt{Signal} is the return type of the function \texttt{getSignalByName}.
Eventually, we establish \textit{depends} relations between the variables in local \texttt{import} statements and the pointing entities in other modules, e.g., the imported \texttt{Signal} points to the class \texttt{Signal} in another module.

\subsection{Dataflow-Guided Retrieval}
\label{sec:retrieval}

Given an unfinished code, we identify fine-grained import information by dataflow analysis and retrieve relevant entities from the repo-specific context graph.
We do not intend to perform precise type inference \cite{Peng2022Static} for a dynamically typed language like Python, but rather provide relevant background knowledge to code LMs, which provides the definitions of code entities such as class members and function arguments.

All cross-file context is indicated by local \texttt{import} statements in Python.
However, only using such coarse-grained import information may overlook the knowledge of its specific usages \cite{Ding2022CoCoMIC}.
We denote import information by $(module, name)$, where $module$ indicates another code file in the repository and $name$ indicates the specific code entity.
Particularly, $name$ can be expanded by its \textit{refers} relations in the extracted DFG.
For example, we would obtain the fine-grained import information $(module,name.attr)$ if there is a code statement containing \texttt{name.attr}.

For each local \texttt{import} statement, we collect a set of fine-grained import information.
Import information points to code entities in the repo-specific context graph, which is achieved through directory structure and string matching. 
Let us see Figure~\ref{fig:framework} for example. 
Given the fine-grained import information (\texttt{RecordSignal}, \texttt{RecordSignal.getSignalByName}), we first identify the corresponding module entity \textit{pyPhasesRecordloader/RecordSignal.py} through directory structure. 
Then, the code entity \texttt{getSignalByName} in class \texttt{RecordSignal} contained in the module is located by string matching.
Finally, the relevant entities are retrieved along \textit{depends} relations using a depth-first search.
The retrieved entities provide comprehensive type-related background knowledge for both cross-file imports and usages in the unfinished code.

\subsection{Prompt Generation}
\label{sec:generation}

Before querying LMs, we restore the retrieved entities to the source code and concatenate it with the unfinished code to generate well-formed prompts.

As the maximum input lengths of LMs are finite and fixed, we employ the dynamic context allocation strategy in \cite{Shrivastava2023Repository}.
It pre-allocates half of the total input lengths for the relevant background knowledge and the other half for the unfinished code.
If either is shorter than the allocated length, the remaining tokens are allocated to the other.
For the overlong unfinished code, we just truncate it to obtain the last tokens.
Given the number $n$ of tokens allocated to the relevant background knowledge, we next describe the generation process of background knowledge.
Refer to Algorithms~\ref{alg:gbk} and~\ref{alg:obk} in Appendix~\ref{appendix:gbk} for details.

Algorithm~\ref{alg:gbk} describes the center control of the generation process.
We consider the entities that have data relations with the line to be completed as the most relevant entities, which are denoted by $E_r$ and form the primary background knowledge.
The entities from other local \texttt{import} statements are denoted by $E_o$ and are incrementally added to the background knowledge until the length exceeds the allocated tokens $n$.
This design helps LMs further understand the context while prevents the primary background knowledge from being truncated.

Our mission is to organize the prompts like the original code to maintain the nature of programs \cite{Hindle2012Naturalness}, as described in Algorithm~\ref{alg:obk}.
We group the relevant entities in modules and merge those with \textit{contains} relations to avoid redundancy, e.g., class members would not be duplicated if the class already exists.
The code entities in the same module are sorted by their starting line numbers.
Therefore, we construct a module graph $G_m$, where an edge $m_1 \rightarrow m_2$ indicates that there exists an entity in $m_1$ that \textit{depends} on an entity in $m_2$.
The modules in prompts are ranked by a pseudo-topological sort with two priorities: 
(\romannumeral1) Dependent modules come first. The content of $m_2$ should be placed in front of that of $m_1$, which is consistent with programming conventions (Lines 10--15 in Algorithm~\ref{alg:obk}).
(\romannumeral2) Once there are multiple options for topological sort in a directed cyclic graph, the relevant modules are placed ahead \cite{Liu2023Lost}.
We prefer the modules that are reachable from the entities in $E_r$ than those in $E_o$ (i.e., relevant or not).
The entities within both $E_r$ and $E_o$ are sorted in ascending order by their starting line numbers, resulting in an ordered list $E_c$ of the import entities (Lines 3--9, 13).
Note that a comment ``\# file path of the module'' is put ahead of each module to indicate the relative directory structure.
Finally, we place the relevant background knowledge inside a multi-line string, which precedes the unfinished code to generate the prompts for querying LMs.

Benefiting from the design of our repo-specific context graph, there are two prompt scopes, named \textit{definition} and \textit{complete}, to control the details of code entities.
Compared with only definitions, prompts under the \textit{complete} scope contain specific function bodies and variable statements.

%====================%
\section{Experiment Setup}

\subsection{Datasets}
The widely-used datasets \cite{Raychev2016Probabilistic, Lu2021CodeXGLUE, Peng2023Revisiting} for code completion only provide a single unfinished code file as input.
Several recent benchmarks \cite{Zhang2023RepoCoder, Liu2023RepoBench} evaluate next-line prediction, which is different from our concern with the currently edited line.
\crosscodeeval \cite{Ding2023CrossCodeEval} is a multilingual benchmark for repository-level code completion, where the statement to be completed has at least one use of cross-file API.
Since we focus on Python, we evaluate our \methodname on the Python subset of \crosscodeeval.

We further build a new Python dataset \recceval with more diverse completion targets. 
See Appendix~\ref{appendix:recceval} for details.
The statistics of \recceval and the Python subset of \crosscodeeval are shown in Table~\ref{tab:stat-datasets}, where the number of tokens is calculated using the StarCoder tokenizer \cite{Li2023Starcoder}.

\subsection{Implementation Details}

We evaluate the retrieval-augmented methods that do not involve training, which excludes several works \cite{Shrivastava2023RepoFusion, Shrivastava2023Repository, lu-etal-2022-reacc}.
See Appendix~\ref{appendix:baseline} for more details:
\begin{compactitem}
\item\textbf{Zero-Shot} directly feeds the unfinished code to code LMs, which evaluates their performance without any cross-file information.

\item\textbf{CCFinder} \cite{Ding2022CoCoMIC} is a cross-file context finder tool retrieving the relevant cross-file context from the pre-built project context graph by \texttt{import} statements.
We conduct experiments for CCFinder-$k$ ($k=1,2$), which indicates that CCFinder retrieves $k$-hop neighbors of cross-file code entities.

\item\textbf{RG-1 and RepoCoder} \cite{Zhang2023RepoCoder} construct a retrieval database through a sliding window and retrieve similar code snippets using text similarity-based retrievers.
RepoCoder is an iterative retrieval-generation framework, which retrieves the database with the results generated in the previous iteration.
RG-1 represents the standard RAG and is the first iteration of RepoCoder.
\end{compactitem}

As shown in Table~\ref{tab:lms}, we conduct comprehensive experiments on seven popular LMs. 
For a method, we first preprocess all repositories in the datasets.
Then, we generate prompts for the unfinished code and record the time used.
Finally, we acquire the completion results by feeding the prompts to LMs.
A prediction is the first line of a completion result.

\begin{table}
\centering
\resizebox{\linewidth}{!}
{
\begin{tabular}{l|cc}
\toprule
Features & \crosscodeeval & \recceval \\
\midrule
\# Repositories              & 471 & 2,635 \\
\# Examples                  & 2,665 & 6,461 \\
Avg. \# files in repository  & 30.5 & 24.6 \\
Avg. \# lines in input       & 73.9 & 113.1 \\
Avg. \# tokens in input      & 938.9 & 1,296.2 \\
\# Last char of input        & dot & any \\
Avg. \# tokens in reference  & 13.2 & 8.6 \\
\bottomrule
\end{tabular}
}
\caption{Statistics of the Python subset of \crosscodeeval and the \recceval dataset that we construct.}
\label{tab:stat-datasets}
\end{table}

\begin{table}
\centering
\resizebox{\linewidth}{!}
{
\begin{tabular}{ll|c}
    \toprule
    \multicolumn{2}{c|}{Models} & Parameter sizes \\
    \midrule
    \multirow{4}{*}{\makecell{Specialized \\models}} & CodeGen    & 350M, 2.7B, 6.1B, 16.1B \\
                                                     & CodeGen2.5 & 7B \\
                                                     & SantaCoder & 1.1B \\
                                                     & StarCoder  & 15.5B \\
    \midrule
    Adapted model & Code Llama & 7B \\
    \midrule
    GPT models & GPT-3.5, GPT-4 & - \\
    \bottomrule
\end{tabular}
}
\caption{The LMs used in our experiments. See Appendix~\ref{appendix:lms} for more details.}
\label{tab:lms}
\end{table}

\begin{table*}
\centering
\resizebox{\linewidth}{!}
{
\begin{tabular}{l|cccc|cccc|cccc|cccc}
    \toprule
    \multirow{3}{*}{Methods} & \multicolumn{4}{c|}{CodeGen-350M} & \multicolumn{4}{c|}{SantaCoder-1.1B} & \multicolumn{4}{c|}{CodeGen25-7B} & \multicolumn{4}{c}{StarCoder-15.5B} \\
    \cmidrule(lr){2-5} \cmidrule(lr){6-9} \cmidrule(lr){10-13} \cmidrule(lr){14-17}
    & EM & ES & ID.EM & F1 & EM & ES & ID.EM & F1 & EM & ES & ID.EM & F1 & EM & ES & ID.EM & F1 \\
    \midrule
    Zero-Shot   & 2.81 & 55.01 & 8.22 & 38.02 & 3.79 & 57.92 & 10.43 & 41.98 & 7.77 & 60.52 & 14.45 & 45.40 & 8.71 & 62.08 & 16.02 & 47.58 \\
    CCFinder-1  & 9.64 & 59.05 & 16.36 & 45.33 & 14.37 & 63.86 & 22.89 & 52.26 & 18.84 & 66.67 & 27.35 & 56.05 & 27.99 & 72.59 & 38.24 & 64.46 \\
    CCFinder-2  & 8.22 & 58.17 & 14.52 & 44.15 & 11.41 & 62.47 & 19.74 & 49.90 & 15.50 & 65.27 & 24.05 & 53.56 & 28.67 & 73.25 & 39.10 & 65.59 \\
    RG-1        & 9.19 & 60.10 & 16.89 & 46.45 & 12.35 & 64.09 & 22.10 & 51.79 & 17.34 & 67.36 & 27.28 & 56.22 & 26.27 & 72.70 & 37.00 & 64.04 \\
    RepoCoder   & 10.13 & 61.25 & 18.65 & 48.29 & 13.62 & 65.53 & 23.94 & 54.06 & 19.51 & 68.98 & 29.57 & 58.51 & 29.12 & 74.56 & 40.83 & 66.81 \\
    \methodname & \textbf{13.02} & \textbf{61.30} & \textbf{20.53} & \textbf{49.04} & \textbf{20.64} & \textbf{67.04} & \textbf{29.83} & \textbf{57.37} & \textbf{24.99} & \textbf{70.10} & \textbf{34.63} & \textbf{61.14} & \textbf{34.67} & \textbf{75.83} & \textbf{45.63} & \textbf{69.93} \\
    \bottomrule
\end{tabular}
}
\caption{Performance comparison on the \crosscodeeval dataset. Numbers are shown in percentage (\%).}
\label{tab:exp-cceval}
\end{table*}

\begin{table*}
\centering
\resizebox{\linewidth}{!}
{
\begin{tabular}{l|cccc|cccc|cccc|cccc}
    \toprule
    \multirow{3}{*}{Methods} & \multicolumn{4}{c|}{CodeGen-350M} & \multicolumn{4}{c|}{SantaCoder-1.1B} & \multicolumn{4}{c|}{CodeGen25-7B} & \multicolumn{4}{c}{StarCoder-15.5B} \\
    \cmidrule(lr){2-5} \cmidrule(lr){6-9} \cmidrule(lr){10-13} \cmidrule(lr){14-17}
    & EM & ES & ID.EM & F1 & EM & ES & ID.EM & F1 & EM & ES & ID.EM & F1 & EM & ES & ID.EM & F1 \\
    \midrule
    Zero-Shot     & \ \ 4.01 & 49.41 & \ \ 9.75 & 25.98 & \ \ 5.54 & 52.95 & 11.93 & 29.94 & 11.10 & 57.25 & 17.37 & 35.55 & 12.77 & 58.84 & 20.03 & 38.12 \\
    CCFinder-1    & 14.15 & 55.75 & 21.24 & 37.74 & 21.36 & 61.90 & 29.31 & 46.18 & 26.87 & 65.76 & 34.55 & 51.00 & 39.33 & 73.05 & 48.18 & 63.49 \\
    CCFinder-2    & 11.64 & 53.70 & 17.94 & 34.15 & 17.12 & 59.57 & 24.58 & 41.93 & 22.49 & 63.42 & 29.72 & 46.81 & 39.92 & 73.29 & 48.91 & 64.08 \\
    RG-1          & 19.44 & 59.08 & 26.02 & 40.92 & 23.62 & 63.23 & 30.58 & 46.24 & 29.33 & 66.94 & 36.06 & 51.36 & 42.67 & 74.64 & 51.11 & 64.64 \\
    RepoCoder     & \textbf{22.46} & \textbf{60.59} & 29.05 & 43.91 & 27.29 & 65.06 & 34.56 & 49.68 & 32.84 & 68.73 & 40.07 & 54.73 & 46.26 & 76.44 & 54.47 & 67.59 \\
    \methodname   & 22.12 & 60.41 & \textbf{29.73} & \textbf{46.09} & \textbf{30.26} & \textbf{66.90} & \textbf{39.08} & \textbf{55.43} & \textbf{36.46} & \textbf{70.76} & \textbf{44.67} & \textbf{60.40} & \textbf{46.49} & \textbf{76.80} & \textbf{55.98} & \textbf{70.32} \\
    \bottomrule
\end{tabular}
}
\caption{Performance comparison on the \recceval dataset.}
\label{tab:exp-recceval}
\end{table*}

\subsection{Evaluation Metrics}

We evaluate the accuracy of each method by code match and identifier match scores \cite{Ding2023CrossCodeEval}, as well as the efficiency by prompt generation time.
We report the average of each metric. See Appendix~\ref{appendix:metric} for more details:
\begin{compactitem}
\item\textbf{Code match.}
We directly compare the prediction with the reference, which is measured using exact match (EM) and edit similarity (ES) \cite{Lu2021CodeXGLUE, Zhang2023RepoCoder}.

\item\textbf{Identifier match.}
We evaluate the predicted APIs by identifier exact match (ID.EM) and F1-score \cite{Ding2023CrossCodeEval}.

\item\textbf{Prompt generation time.}
We record the prompt generation time to evaluate the efficiency of each method, which is a new and significant metric for real-time completion.
\end{compactitem}

%====================%
\section{Experimental Results and Analysis}

\subsection{Performance Comparison}

The performance comparison on the \crosscodeeval and \recceval datasets is listed in Tables~\ref{tab:exp-cceval} and~\ref{tab:exp-recceval}, respectively.
Additional results on other CodeGen models are supplemented in Appendix~\ref{appendix:performance}.
Overall, \methodname significantly improves the accuracy of various code LMs. 
Particularly, the CodeGen-350M model integrated with \methodname even outperforms the zero-shot StarCoder-15.5B model.

In comparison to other retrieval-augmented methods, \methodname also shows generally superior accuracy across all settings.
The average absolute improvement on EM, ES, ID.EM, and F1 versus RepoCoder is 3.43\%, 1.00\%, 3.62\%, and 3.27\%, respectively.
RepoCoder retrieves similar code demonstrations that help increase the ES metric of completion results.
But RepoCoder ignores the validity of its generated identifiers in private repositories, which decreases the metrics for code exact match and identifier match.
Such almost correct completion results may introduce unconscious bugs to the programmers who are unfamiliar with the repository.
In contrast, \methodname presents the definitions of relevant code entities, providing better control over code LMs to generate valid identifiers.
Moreover, the background knowledge can be used as a reference to help the programmers understand and review the completion results in IDEs.
\methodname using the CodeGen-350M model is slightly worse than RepoCoder in terms of code match metrics on the \recceval dataset, as the model may not be powerful enough to capture the data relations in our provided background knowledge.

CCFinder retrieves cross-file code entities by plain import relations.
The entities retrieved by CCFinder were originally designed to be encoded for training code LMs.
When used as a retrieval-augmented method, CCFinder retrieves too many code entities based on coarse-grained import information, resulting in truncation of truly relevant context.
As a result, CCFinder-2 with more retrieved entities only exceeds CCFinder-1 slightly on the StarCoder model that supports longer inputs.
Therefore, the subsequent analysis experiments are conducted with CCFinder-1.
Guiding by our dataflow analysis, \methodname retrieves relevant code entities more precisely, leading to significantly superior accuracy.

The performance of code completion varies on the two datasets.
First, according to the statistics in Table~\ref{tab:stat-datasets}, the average reference length of \recceval is significantly shorter than that of \crosscodeeval, leading to the higher EM metrics of both code and identifier on \recceval.
Moreover, all inputs of \crosscodeeval end with a dot where a correct API is required in the first place, which is more suitable for CCFinder and \methodname that retrieve code definitions.
Many inputs of \recceval end with partial names of the target APIs, which facilitates text similarity-based retrievals including RG-1 and RepoCoder.
Therefore, the lead of \methodname on \crosscodeeval is more significant.

\subsection{Efficiency Evaluation}

The time spent on prompt generation is perceived by users whenever code completion is triggered.
Table~\ref{tab:time} shows the prompt generation time of each method using the CodeGen-350M model, and additional results are shown in Appendix~\ref{appendix:time}.
CCFinder and \methodname require parsing the unfinished code into an AST or a DFG, which is slightly slower than RG-1 with text similarity-based retrieval but still comparable.
RepoCoder relies on RG-1 to generate sufficient content for the second retrieval, which results in more than 4 seconds even on the smallest CodeGen-350M model and may not be feasible for real-time code completion.

In summary, \methodname is efficient for real-time code completion in IDEs. 
Compared to the methods with comparable efficiency (i.e., excluding RepoCoder), \methodname is considerably ahead in the accuracy of repository-level code completion.

% Time
\begin{table}
\centering
{\small
\begin{tabular}{l|cc}
    \toprule
    Methods & \crosscodeeval & \recceval \\
    \midrule
    CCFinder-1    & 32 & 49 \\
    CCFinder-2    & 52 & 82 \\
    RG-1          & \textbf{13} & \textbf{15} \\
    RepoCoder     & 4,062 & 4,413 \\  
    \methodname   & 40 & 44 \\
    \bottomrule
\end{tabular}}
\caption{Prompt generation time (in milliseconds) of each method using the CodeGen-350M model.}
\label{tab:time}
\end{table}

% Ablation experiments
\begin{table*}
\centering
\resizebox{\linewidth}{!}
{
\begin{tabular}{l|cccc|cccc|cccc|cccc}
    \toprule
    \multirow{3}{*}{Methods} & \multicolumn{4}{c|}{CodeGen-350M} & \multicolumn{4}{c|}{SantaCoder-1.1B} & \multicolumn{4}{c|}{CodeGen25-7B} & \multicolumn{4}{c}{StarCoder-15.5B} \\
    \cmidrule(lr){2-5} \cmidrule(lr){6-9} \cmidrule(lr){10-13} \cmidrule(lr){14-17}
    & EM & ES & ID.EM & F1 & EM & ES & ID.EM & F1 & EM & ES & ID.EM & F1 & EM & ES & ID.EM & F1 \\
    
    \midrule
    \methodname            & \textbf{13.02} & \textbf{61.30} & \textbf{20.53} & \textbf{49.04} & \textbf{20.64} & \textbf{67.04} & \textbf{29.83} & \textbf{57.37} & \textbf{24.99} & \textbf{70.10} & \textbf{34.63} & \textbf{61.14} & \textbf{34.67} & \textbf{75.83} & \textbf{45.63} & \textbf{69.93} \\

    \quad w/o cross\_df    & 12.12 & 60.93 & 19.51 & 48.32 & 18.42 & 66.05 & 27.62 & 55.64 & 22.59 & 69.15 & 31.89 & 59.36 & 30.73 & 73.85 & 41.05 & 66.31 \\
    \quad w/o intra\_df    & 10.88 & 59.74 & 17.56 & 46.25 & 15.95 & 64.11 & 24.09 & 52.72 & 19.59 & 67.08 & 28.33 & 56.14 & 32.35 & 74.60 & 43.00 & 67.98 \\
    \quad w/o dataflow    & 10.13 & 59.55 & 17.00 & 45.88 & 14.90 & 63.57 & 23.11 & 51.88 & 18.57 & 66.85 & 27.13 & 55.53 & 28.82 & 72.80 & 38.87 & 64.65 \\
    
    \bottomrule
\end{tabular}
}
\caption{Ablation study for dataflow analysis on the \crosscodeeval dataset.}
\label{tab:abl-cceval}
\end{table*}

\begin{table*}
\centering
\resizebox{\linewidth}{!}
{
\begin{tabular}{l|ccccc|ccccc|cccc}
    \toprule
    \multirow{3}{*}{Methods} & \multicolumn{5}{c|}{GPT-3.5} & \multicolumn{5}{c|}{GPT-4} & \multicolumn{4}{c}{StarCoder-15.5B} \\
    \cmidrule(lr){2-6} \cmidrule(lr){7-11} \cmidrule(lr){12-15} 
    & EM & ES & ID.EM & F1 & WOF & EM & ES & ID.EM & F1 & WOF & EM & ES & ID.EM & F1 \\

    \midrule
    Zero-Shot   & 10.00 & 57.20 & 10.00 & 44.39 & 0 & 16.00 & 67.20 & 20.00 & 56.82 & 0 & 12.00 & 65.44 & 18.00 & 53.18 \\
    CCFinder-1  & 20.00 & 66.32 & \textbf{30.00} & 57.12 & 0 & 38.00 & 74.80 & 46.00 & 69.36 & 4 & 34.00 & 76.24 & 46.00 & 72.31 \\
    RG-1        & 14.00 & 54.06 & 20.00 & 43.06 & 9 & 12.00 & 35.20 & 18.00 & 22.30 & 34 & 20.00 & 74.36 & 38.00 & 70.08 \\
    RepoCoder   & 18.00 & 63.44 & 22.00 & 57.13 & 1 & 34.00 & 73.18 & 40.00 & 65.07 & 7 & 26.00 & 72.48 & 42.00 & 68.99 \\
    \methodname & \textbf{24.00} & \textbf{67.54} & \textbf{30.00} & \textbf{58.08} & 0 & \textbf{42.00} & \textbf{76.58} & \textbf{50.00} & \textbf{72.36} & 5 & \textbf{38.00} & \textbf{77.84} & \textbf{52.00} & \textbf{77.73} \\
    \bottomrule
\end{tabular}
}
\caption{Performance comparison on the sampled \crosscodeeval dataset. ``WOF'' is a manual count indicating the number of predictions with wrong output format, such as ``The last line of the code should be:''.}
\label{tab:exp-gpt}
\end{table*}

\subsection{Ablation Study}
\label{sec:abl}

To analyze the effectiveness of dataflow analysis in \methodname, we conduct an ablation study shown in Tables~\ref{tab:abl-cceval} and~\ref{tab:abl-recceval}.
``w/o cross\_df'' disables \textit{depends} relations in the repo-specific context graph, making \methodname unable to handle the data dependency relations in other code files.
``w/o intra\_df'' disables the dataflow analysis for the unfinished code, which only allows \methodname to retrieve coarse-grained import information in the order of their starting line numbers.
``w/o dataflow'' degenerates \methodname into a naive method that simply takes the imported cross-file entities in the unfinished code as the relevant background knowledge.

The ablation study demonstrates that the complete \methodname achieves the best performance, and all usages of dataflow analysis play a positive role in repository-level code completion.
It can be observed that the enhancement of the ``intra\_df'' component on the StarCoder model is less than that on other models.
This component places the more relevant background knowledge in front of the prompt to prevent truncation, which is weakened to some extent on the StarCoder model with a maximum context length of 8K tokens.

The accuracy of \methodname without dataflow analysis is still comparable with CCFinder.
CCFinder groups the relevant context in code entities, which is counter-intuitive for source code (see the example shown in Appendix~\ref{appendix:prompt}).
The results reveal that the well-formed prompts generated by \methodname can better steer code LMs, even if the depth-first search for dependent code entities is absent.

% Prompt scopes
\begin{figure}
    \centering
    \includegraphics[width=\linewidth]{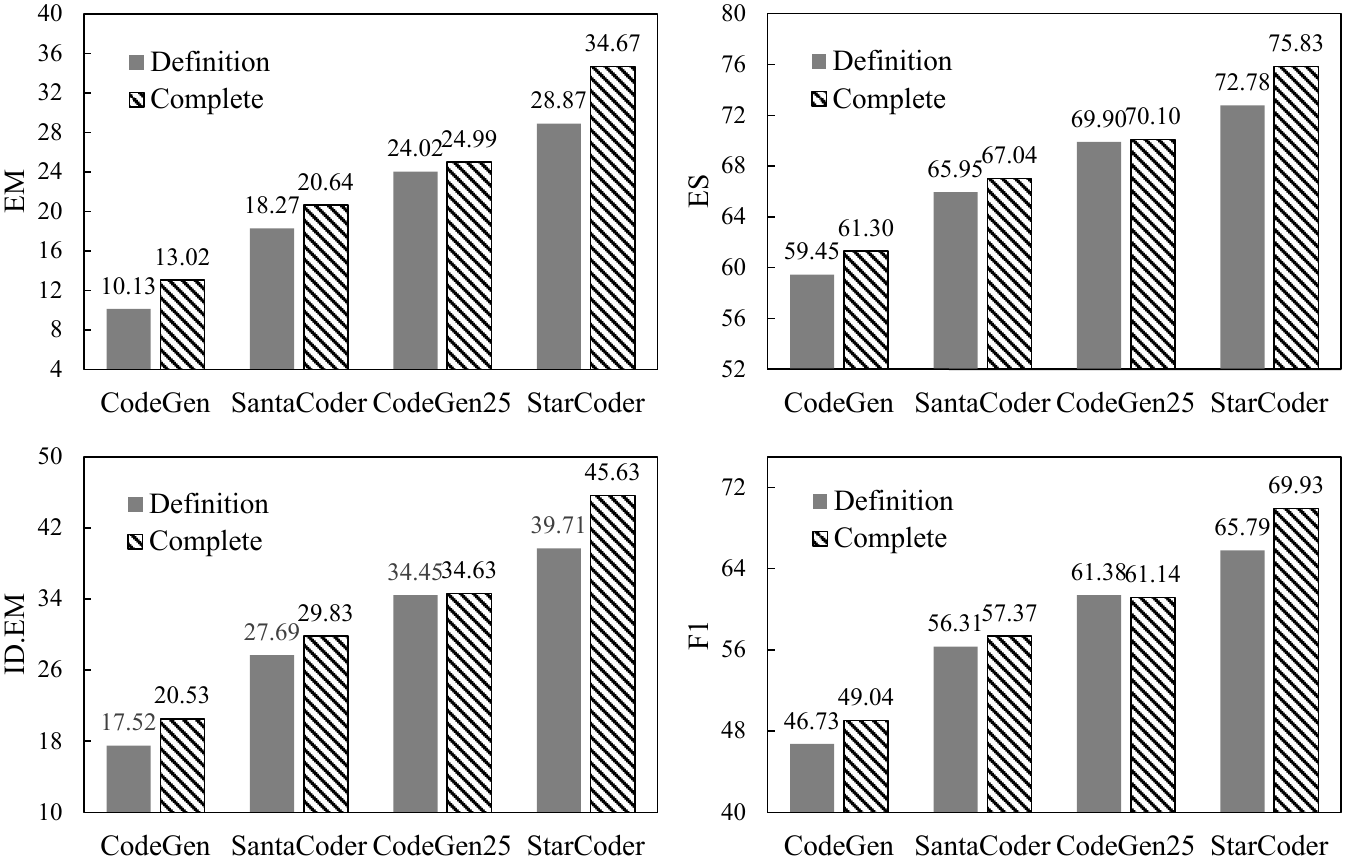}
    \caption{Performance comparison of two prompt scopes on the \crosscodeeval dataset.}
    \label{fig:scope-cceval}
\end{figure}

\subsection{Analysis of Prompt Scopes}

The prompts generated by \methodname consist of the definitions of code entities, which provide options for the \textit{definition} and \textit{complete} scopes, as described in Section~\ref{sec:generation}.
We further conduct experiments to evaluate the influence of the two prompt scopes.
The results on the \crosscodeeval and \recceval datasets are shown in Figures~\ref{fig:scope-cceval} and~\ref{fig:scope-recc}, respectively.

\methodname with the \textit{complete} scope achieves the best performance across all settings, which indicates that code implementation can further enhance LMs. 
Implementation details can provide a deeper understanding of code entities, along with the programming styles.
Moreover, \methodname with the \textit{definition} scope outperforms CCFinder and RG-1 in most settings (cf. Tables~\ref{tab:exp-cceval} and~\ref{tab:exp-recceval}), suggesting that the definitions without specific implementations are also useful for code LMs.
Since an implementation is usually much longer than its definitions, both prompt scopes are optional in practical applications, in a trade-off between accuracy and cost.

\subsection{Analysis of Adapted LM}
\label{sec:adlm}

\begin{figure}
    \centering
    \includegraphics[width=\linewidth]{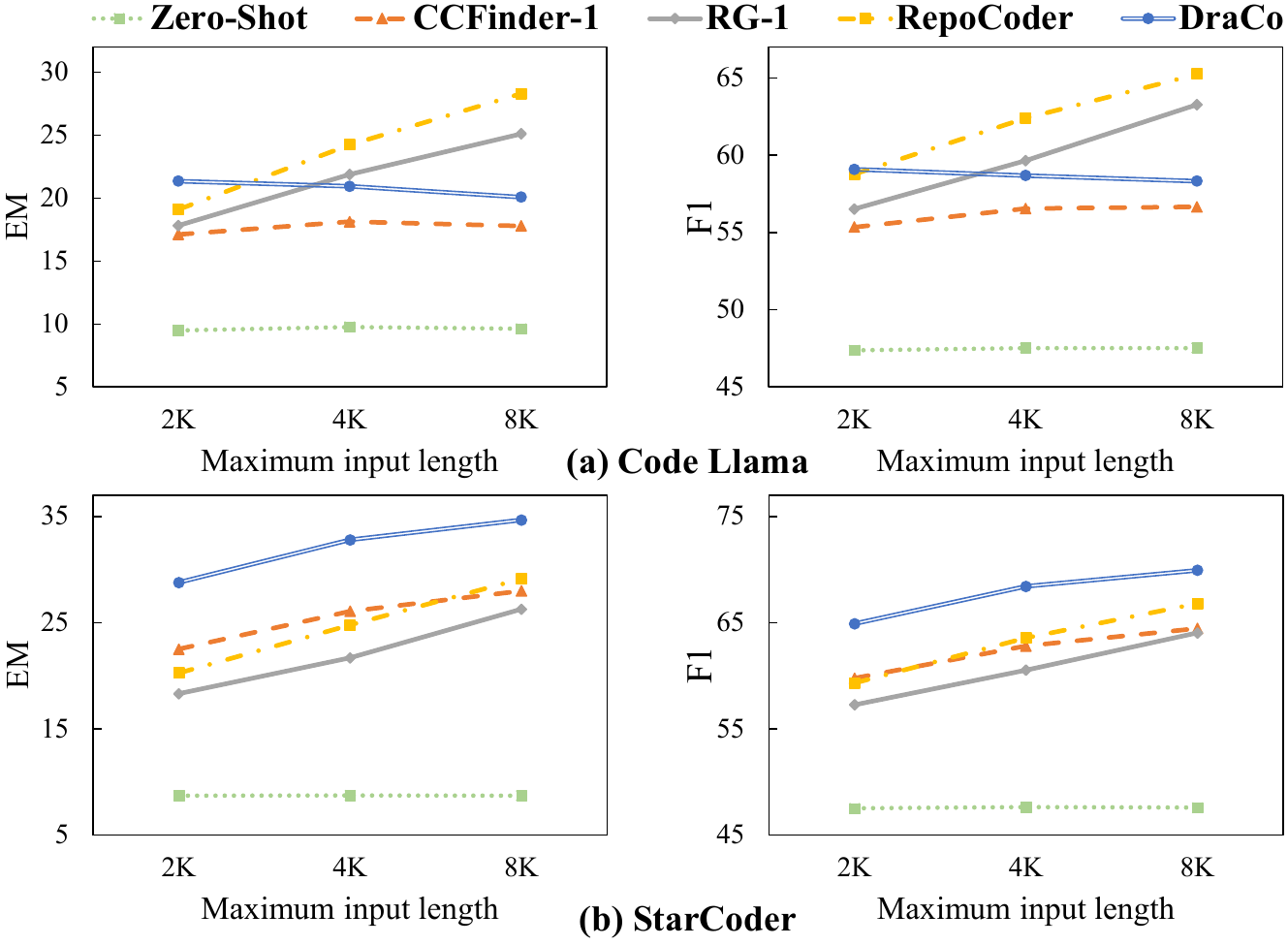}
    \caption{Performance changes with different maximum input lengths on the \crosscodeeval dataset.}
    \label{fig:input-exps}
\end{figure}

We analyze the effect of different maximum input lengths for different types of LMs.
Specifically, we evaluate Code Llama-7B and StarCoder-15.5B with 2K, 4K, and 8K tokens.
The performance changes are shown in Figure~\ref{fig:input-exps}, and the complete results are presented in Tables~\ref{tab:exp-codellama-cceval}--\ref{tab:exp-starcoder-recceval}.

With the increase of the maximum input length, the accuracy of \methodname applied to Code Llama decreases, which shows the opposite trend of the StarCoder model.
Code Llama is created by further training Llama 2 on its code-specific datasets. 
It is different from specialized code LMs such as StarCoder which are mainly pre-trained on code corpora.
With similar background knowledge, Code Llama-7B does not have enough capability to capture data dependency relations in long Python code, leading to the degraded accuracy of \methodname.
In contrast, specialized code LMs can understand longer code context and may be a better choice for repository-level code completion.

\subsection{Analysis of GPT Models}

We randomly sample 50 examples from \crosscodeeval and evaluate them with GPT-3.5, GPT-4, and StarCoder-15.5B.
The results shown in Table~\ref{tab:exp-gpt} reveal that: 
(\romannumeral1) \methodname can also significantly enhance GPT models and achieve superior accuracy. 
(\romannumeral2) Code completion with GPT models suffers from the difficulties in output format and API cost. 
Given the long context of repository-level code completion, there lacks sufficient length to place the demonstrations required for in-context learning \cite{Brown2020Language}. 
It is hard to control the output format of GPT models through instruction, which may introduce bias into the evaluation, especially for RG-1 with GPT-4. 
Moreover, the API cost for this evaluation (only 50 examples) is nearly 35 US dollars.
(\romannumeral3) Excluding the effect of wrong output format, we may assume ``GPT-4 > StarCoder-15.5B > GPT-3.5'' for this task.

%====================%
\section{Conclusions}

This paper proposes \methodname, a dataflow-guided retrieval augmentation approach for repository-level code completion.
To guide more precise retrieval, an extended dataflow analysis is designed by setting type-sensitive data dependency relations.
\methodname indexes private repositories to form repo-specific context graphs and retrieves relevant background knowledge from them, which is assembled with the unfinished code to generate well-formed prompts for querying code LMs.
The experiments on the \crosscodeeval dataset and our \recceval dataset demonstrate the superior accuracy and applicable efficiency of \methodname.
In future work, we will explore other structured code representations.

%====================%
\section*{Acknowledgments}
We thank the anonymous reviewers for their valuable comments.
This work was supported by the National Natural Science Foundation of China (No. 62272219) and the Collaborative Innovation Center of Novel Software Technology \& Industrialization.

%====================%

\section*{Ethical Considerations}

The code generated by pre-trained LMs may contain non-existent APIs or even introduce potential bugs.
The retrieval-augmented approaches including ours mitigate this issue only to some extent.
We recommend presenting our retrieved background knowledge to programmers for review and taking appropriate care of these risks if deploying our approach in real-world applications.

All the datasets and code LMs used in this work are publicly available with permissive licenses.
The \crosscodeeval dataset and CodeGen family are licensed under the Apache-2.0 License.
The SantaCoder and StarCoder models are licensed under the BigCode OpenRAIL-M v1 license agreement.
Code Llama is governed by the Meta license.\footnote{\url{https://github.com/facebookresearch/llama/blob/main/LICENSE}}
The repositories in our \recceval dataset are all licensed under permissive licenses including MIT, Apache, and BSD licenses.

\section*{Limitations}

\methodname relies on a code LM to support long inputs and capture data dependency relations in the provided background knowledge.
Thus, the performance of \methodname may be limited by the capability of the code LM.
According to our experiments, \methodname still has a considerable improvement on the smallest CodeGen-350M model with 2K tokens, which mitigates this limitation.

The effectiveness of \methodname may degrade when the code intent is unclear.
For new line or function body completion, the guidance of dataflow analysis is weakened since \methodname may not be able to set priorities for import information.
We focus on the code completion of currently edited line and evaluate multi-line code completion in Appendix~\ref{appendix:multi}.
Future work may explore the role of dataflow analysis in different completion scenarios.

\methodname requires changes to migrate to other programming languages.
Our idea of guiding retrieval with dataflow analysis is not limited to Python.
However, due to the different characteristics of programming languages, \methodname needs to extend dataflow analysis for target languages.
The variety of static analysis tools for common programming languages provides convenience for implementing multilingual \methodname.

\bibliography{anthology,custom}

\begin{thebibliography}{66}
\expandafter\ifx\csname natexlab\endcsname\relax\def\natexlab#1{#1}\fi

\bibitem[{Allal et~al.(2023)Allal, Li, Kocetkov, Mou, Akiki, Ferrandis, Muennighoff, Mishra, Gu, Dey, Umapathi, Anderson, Zi, Lamy{-}Poirier, Schoelkopf, Troshin, Abulkhanov, Romero, Lappert, Toni, del R{\'{\i}}o, Liu, Bose, Bhattacharyya, Zhuo, Yu, Villegas, Zocca, Mangrulkar, Lansky, Nguyen, Contractor, Villa, Li, Bahdanau, Jernite, Hughes, Fried, Guha, de~Vries, and von Werra}]{Allal2023Santacoder}
Loubna~Ben Allal, Raymond Li, Denis Kocetkov, Chenghao Mou, Christopher Akiki, Carlos~Mu{\~{n}}oz Ferrandis, Niklas Muennighoff, Mayank Mishra, Alex Gu, Manan Dey, Logesh~Kumar Umapathi, Carolyn~Jane Anderson, Yangtian Zi, Joel Lamy{-}Poirier, Hailey Schoelkopf, Sergey Troshin, Dmitry Abulkhanov, Manuel Romero, Michael Lappert, Francesco~De Toni, Bernardo~Garc{\'{\i}}a del R{\'{\i}}o, Qian Liu, Shamik Bose, Urvashi Bhattacharyya, Terry~Yue Zhuo, Ian Yu, Paulo Villegas, Marco Zocca, Sourab Mangrulkar, David Lansky, Huu Nguyen, Danish Contractor, Luis Villa, Jia Li, Dzmitry Bahdanau, Yacine Jernite, Sean Hughes, Daniel Fried, Arjun Guha, Harm de~Vries, and Leandro von Werra. 2023.
\newblock {SantaCoder}: don't reach for the stars!
\newblock \emph{CoRR}, 2301.03988:1--35.

\bibitem[{Barke et~al.(2023)Barke, James, and Polikarpova}]{Barke2023Grounded}
Shraddha Barke, Michael~B. James, and Nadia Polikarpova. 2023.
\newblock Grounded {Copilot}: How programmers interact with code-generating models.
\newblock \emph{Proc. {ACM} Program. Lang.}, 7({OOPSLA1}):85--111.

\bibitem[{Barnett and Constantine(1968)}]{Barnett1968Modular}
Tom~O. Barnett and Larry~L. Constantine. 1968.
\newblock \emph{Modular Programming: Proceedings of a National Symposium}.
\newblock Information \& Systems Institute, Leipzig, Germany.

\bibitem[{Brown et~al.(2020)Brown, Mann, Ryder, Subbiah, Kaplan, Dhariwal, Neelakantan, Shyam, Sastry, Askell, Agarwal, Herbert{-}Voss, Krueger, Henighan, Child, Ramesh, Ziegler, Wu, Winter, Hesse, Chen, Sigler, Litwin, Gray, Chess, Clark, Berner, McCandlish, Radford, Sutskever, and Amodei}]{Brown2020Language}
Tom~B. Brown, Benjamin Mann, Nick Ryder, Melanie Subbiah, Jared Kaplan, Prafulla Dhariwal, Arvind Neelakantan, Pranav Shyam, Girish Sastry, Amanda Askell, Sandhini Agarwal, Ariel Herbert{-}Voss, Gretchen Krueger, Tom Henighan, Rewon Child, Aditya Ramesh, Daniel~M. Ziegler, Jeffrey Wu, Clemens Winter, Christopher Hesse, Mark Chen, Eric Sigler, Mateusz Litwin, Scott Gray, Benjamin Chess, Jack Clark, Christopher Berner, Sam McCandlish, Alec Radford, Ilya Sutskever, and Dario Amodei. 2020.
\newblock Language models are few-shot learners.
\newblock In \emph{NeurIPS}, pages 1877--1901, Virtual.

\bibitem[{Cai et~al.(2022)Cai, Wang, Liu, and Shi}]{Cai2022Recent}
Deng Cai, Yan Wang, Lemao Liu, and Shuming Shi. 2022.
\newblock Recent advances in retrieval-augmented text generation.
\newblock In \emph{SIGIR}, pages 3417--3419, Madrid, Spain. ACM.

\bibitem[{Chen et~al.(2021)Chen, Tworek, Jun, Yuan, de~Oliveira~Pinto, Kaplan, Edwards, Burda, Joseph, Brockman, Ray, Puri, Krueger, Petrov, Khlaaf, Sastry, Mishkin, Chan, Gray, Ryder, Pavlov, Power, Kaiser, Bavarian, Winter, Tillet, Such, Cummings, Plappert, Chantzis, Barnes, Herbert-Voss, Guss, Nichol, Paino, Tezak, Tang, Babuschkin, Balaji, Jain, Saunders, Hesse, Carr, Leike, Achiam, Misra, Morikawa, Radford, Knight, Brundage, Murati, Mayer, Welinder, McGrew, Amodei, McCandlish, Sutskever, and Zaremba}]{Chen2021Codex}
Mark Chen, Jerry Tworek, Heewoo Jun, Qiming Yuan, Henrique~Ponde de~Oliveira~Pinto, Jared Kaplan, Harri Edwards, Yuri Burda, Nicholas Joseph, Greg Brockman, Alex Ray, Raul Puri, Gretchen Krueger, Michael Petrov, Heidy Khlaaf, Girish Sastry, Pamela Mishkin, Brooke Chan, Scott Gray, Nick Ryder, Mikhail Pavlov, Alethea Power, Lukasz Kaiser, Mohammad Bavarian, Clemens Winter, Philippe Tillet, Felipe~Petroski Such, Dave Cummings, Matthias Plappert, Fotios Chantzis, Elizabeth Barnes, Ariel Herbert-Voss, William~Hebgen Guss, Alex Nichol, Alex Paino, Nikolas Tezak, Jie Tang, Igor Babuschkin, Suchir Balaji, Shantanu Jain, William Saunders, Christopher Hesse, Andrew~N. Carr, Jan Leike, Josh Achiam, Vedant Misra, Evan Morikawa, Alec Radford, Matthew Knight, Miles Brundage, Mira Murati, Katie Mayer, Peter Welinder, Bob McGrew, Dario Amodei, Sam McCandlish, Ilya Sutskever, and Wojciech Zaremba. 2021.
\newblock Evaluating large language models trained on code.
\newblock \emph{CoRR}, 2107.03374:1--35.

\bibitem[{Dakhel et~al.(2023)Dakhel, Majdinasab, Nikanjam, Khomh, Desmarais, and Jiang}]{Dakhel2023GitHub}
Arghavan~Moradi Dakhel, Vahid Majdinasab, Amin Nikanjam, Foutse Khomh, Michel~C. Desmarais, and Zhen Ming~(Jack) Jiang. 2023.
\newblock {GitHub} {Copilot} {AI} pair programmer: Asset or liability?
\newblock \emph{J. Syst. Softw.}, 203:111734.

\bibitem[{Ding et~al.(2023)Ding, Wang, Ahmad, Ding, Tan, Jain, Ramanathan, Nallapati, Bhatia, Roth, and Xiang}]{Ding2023CrossCodeEval}
Yangruibo Ding, Zijian Wang, Wasi~Uddin Ahmad, Hantian Ding, Ming Tan, Nihal Jain, Murali~Krishna Ramanathan, Ramesh Nallapati, Parminder Bhatia, Dan Roth, and Bing Xiang. 2023.
\newblock {CrossCodeEval}: A diverse and multilingual benchmark for cross-file code completion.
\newblock In \emph{NeurIPS}, pages 1--23, New Orleans, LA, USA.

\bibitem[{Ding et~al.(2024)Ding, Wang, Ahmad, Ramanathan, Nallapati, Bhatia, Roth, and Xiang}]{Ding2022CoCoMIC}
Yangruibo Ding, Zijian Wang, Wasi~Uddin Ahmad, Murali~Krishna Ramanathan, Ramesh Nallapati, Parminder Bhatia, Dan Roth, and Bing Xiang. 2024.
\newblock {CoCoMIC}: Code completion by jointly modeling in-file and cross-file context.
\newblock In \emph{COLING}, pages 3433--3445, Torino, Italy. {ELRA} and {ICCL}.

\bibitem[{Feng et~al.(2023)Feng, Ma, Yu, Huang, Wang, Chen, Peng, Feng, Qin, and Liu}]{Feng2023Trends}
Zhangyin Feng, Weitao Ma, Weijiang Yu, Lei Huang, Haotian Wang, Qianglong Chen, Weihua Peng, Xiaocheng Feng, Bing Qin, and Ting Liu. 2023.
\newblock Trends in integration of knowledge and large language models: A survey and taxonomy of methods, benchmarks, and applications.
\newblock \emph{CoRR}, 2311.05876:1--22.

\bibitem[{Hashimoto et~al.(2018)Hashimoto, Guu, Oren, and Liang}]{Hashimoto2018Retrieve}
Tatsunori~B. Hashimoto, Kelvin Guu, Yonatan Oren, and Percy Liang. 2018.
\newblock A retrieve-and-edit framework for predicting structured outputs.
\newblock In \emph{NeurIPS}, pages 10073--10083, Montr{\'{e}}al, Canada.

\bibitem[{He et~al.(2021)He, Xu, Zhang, Hao, Feng, and Xu}]{He2021PyART}
Xincheng He, Lei Xu, Xiangyu Zhang, Rui Hao, Yang Feng, and Baowen Xu. 2021.
\newblock {PyART}: Python {API} recommendation in real-time.
\newblock In \emph{ICSE}, pages 1634--1645, Madrid, Spain. IEEE.

\bibitem[{Hindle et~al.(2012)Hindle, Barr, Su, Gabel, and Devanbu}]{Hindle2012Naturalness}
Abram Hindle, Earl~T. Barr, Zhendong Su, Mark Gabel, and Premkumar~T. Devanbu. 2012.
\newblock On the naturalness of software.
\newblock In \emph{ICSE}, pages 837--847, Zurich, Switzerland. IEEE.

\bibitem[{Husain et~al.(2019)Husain, Wu, Gazit, Allamanis, and Brockschmidt}]{husain2019codesearchnet}
Hamel Husain, Ho{-}Hsiang Wu, Tiferet Gazit, Miltiadis Allamanis, and Marc Brockschmidt. 2019.
\newblock {CodeSearchNet} challenge: Evaluating the state of semantic code search.
\newblock \emph{CoRR}, 1909.09436:1--6.

\bibitem[{Izadi et~al.(2022)Izadi, Gismondi, and Gousios}]{Izadi2022CodeFill}
Maliheh Izadi, Roberta Gismondi, and Georgios Gousios. 2022.
\newblock {CodeFill}: Multi-token code completion by jointly learning from structure and naming sequences.
\newblock In \emph{ICSE}, pages 401--412, Pittsburgh, PA, USA. ACM.

\bibitem[{Karpukhin et~al.(2020)Karpukhin, Oguz, Min, Lewis, Wu, Edunov, Chen, and Yih}]{karpukhin-etal-2020-dense}
Vladimir Karpukhin, Barlas Oguz, Sewon Min, Patrick Lewis, Ledell Wu, Sergey Edunov, Danqi Chen, and Wen-tau Yih. 2020.
\newblock \href {https://doi.org/10.18653/v1/2020.emnlp-main.550} {Dense passage retrieval for open-domain question answering}.
\newblock In \emph{Proceedings of the 2020 Conference on Empirical Methods in Natural Language Processing (EMNLP)}, pages 6769--6781, Online. Association for Computational Linguistics.

\bibitem[{Kazemitabaar et~al.(2023)Kazemitabaar, Chow, Ma, Ericson, Weintrop, and Grossman}]{Kazemitabaar2023Studying}
Majeed Kazemitabaar, Justin Chow, Carl Ka~To Ma, Barbara~J. Ericson, David Weintrop, and Tovi Grossman. 2023.
\newblock Studying the effect of {AI} code generators on supporting novice learners in introductory programming.
\newblock In \emph{CHI}, pages 455:1--455:23, Hamburg, Germany. ACM.

\bibitem[{Kim et~al.(2021)Kim, Zhao, Tian, and Chandra}]{Kim2021Code}
Seohyun Kim, Jinman Zhao, Yuchi Tian, and Satish Chandra. 2021.
\newblock Code prediction by feeding trees to transformers.
\newblock In \emph{ICSE}, pages 150--162, Madrid, Spain. IEEE.

\bibitem[{Kocetkov et~al.(2023)Kocetkov, Li, allal, LI, Mou, Jernite, Mitchell, Ferrandis, Hughes, Wolf, Bahdanau, Werra, and de~Vries}]{Kocetkov2022Stack}
Denis Kocetkov, Raymond Li, Loubna~Ben allal, Jia LI, Chenghao Mou, Yacine Jernite, Margaret Mitchell, Carlos~Mu{\~n}oz Ferrandis, Sean Hughes, Thomas Wolf, Dzmitry Bahdanau, Leandro~Von Werra, and Harm de~Vries. 2023.
\newblock \href {https://openreview.net/forum?id=pxpbTdUEpD} {The {Stack}: 3 {TB} of permissively licensed source code}.
\newblock \emph{TMLR}.

\bibitem[{Le et~al.(2022)Le, Wang, Gotmare, Savarese, and Hoi}]{Le2022CodeRL}
Hung Le, Yue Wang, Akhilesh~Deepak Gotmare, Silvio Savarese, and Steven~Chu{-}Hong Hoi. 2022.
\newblock {CodeRL}: Mastering code generation through pretrained models and deep reinforcement learning.
\newblock In \emph{NeurIPS}, pages 1--15, New Orleans, LA, USA. Curran Associates Inc.

\bibitem[{Levine et~al.(2022)Levine, Dalmedigos, Ram, Zeldes, Jannai, Muhlgay, Osin, Lieber, Lenz, Shalev{-}Shwartz, Shashua, Leyton{-}Brown, and Shoham}]{Levine2022Standing}
Yoav Levine, Itay Dalmedigos, Ori Ram, Yoel Zeldes, Daniel Jannai, Dor Muhlgay, Yoni Osin, Opher Lieber, Barak Lenz, Shai Shalev{-}Shwartz, Amnon Shashua, Kevin Leyton{-}Brown, and Yoav Shoham. 2022.
\newblock Standing on the shoulders of giant frozen language models.
\newblock \emph{CoRR}, 2204.10019:1--19.

\bibitem[{Lewis et~al.(2020)Lewis, Perez, Piktus, Petroni, Karpukhin, Goyal, K{\"{u}}ttler, Lewis, Yih, Rockt{\"{a}}schel, Riedel, and Kiela}]{Lewis2020Retrieval}
Patrick S.~H. Lewis, Ethan Perez, Aleksandra Piktus, Fabio Petroni, Vladimir Karpukhin, Naman Goyal, Heinrich K{\"{u}}ttler, Mike Lewis, Wen{-}tau Yih, Tim Rockt{\"{a}}schel, Sebastian Riedel, and Douwe Kiela. 2020.
\newblock Retrieval-augmented generation for knowledge-intensive {NLP} tasks.
\newblock In \emph{NeurIPS}, pages 9459--9474, Virtual.

\bibitem[{Li et~al.(2024)Li, Li, Zhang, Dong, and Jin}]{li2024evocodebench}
Jia Li, Ge~Li, Xuanming Zhang, Yihong Dong, and Zhi Jin. 2024.
\newblock {EvoCodeBench}: An evolving code generation benchmark aligned with real-world code repositories.
\newblock \emph{CoRR}, 2404.00599:1--15.

\bibitem[{Li et~al.(2023{\natexlab{a}})Li, Zhao, Li, Li, and Jin}]{Li2023Towards}
Jia Li, Yunfei Zhao, Yongmin Li, Ge~Li, and Zhi Jin. 2023{\natexlab{a}}.
\newblock {AceCoder}: Utilizing existing code to enhance code generation.
\newblock \emph{CoRR}, 2303.17780:1--12.

\bibitem[{Li et~al.(2018)Li, Wang, Lyu, and King}]{Li2018Code}
Jian Li, Yue Wang, Michael~R. Lyu, and Irwin King. 2018.
\newblock Code completion with neural attention and pointer networks.
\newblock In \emph{IJCAI}, pages 4159--4165, Stockholm, Sweden. ijcai.org.

\bibitem[{Li et~al.(2023{\natexlab{b}})Li, allal, Zi, Muennighoff, Kocetkov, Mou, Marone, Akiki, LI, Chim, Liu, Zheltonozhskii, Zhuo, Wang, Dehaene, Lamy-Poirier, Monteiro, Gontier, Yee, Umapathi, Zhu, Lipkin, Oblokulov, Wang, Murthy, Stillerman, Patel, Abulkhanov, Zocca, Dey, Zhang, Bhattacharyya, Yu, Luccioni, Villegas, Zhdanov, Lee, Timor, Ding, Schlesinger, Schoelkopf, Ebert, Dao, Mishra, Gu, Anderson, Dolan-Gavitt, Contractor, Reddy, Fried, Bahdanau, Jernite, Ferrandis, Hughes, Wolf, Guha, Werra, and de~Vries}]{Li2023Starcoder}
Raymond Li, Loubna~Ben allal, Yangtian Zi, Niklas Muennighoff, Denis Kocetkov, Chenghao Mou, Marc Marone, Christopher Akiki, Jia LI, Jenny Chim, Qian Liu, Evgenii Zheltonozhskii, Terry~Yue Zhuo, Thomas Wang, Olivier Dehaene, Joel Lamy-Poirier, Joao Monteiro, Nicolas Gontier, Ming-Ho Yee, Logesh~Kumar Umapathi, Jian Zhu, Ben Lipkin, Muhtasham Oblokulov, Zhiruo Wang, Rudra Murthy, Jason~T Stillerman, Siva~Sankalp Patel, Dmitry Abulkhanov, Marco Zocca, Manan Dey, Zhihan Zhang, Urvashi Bhattacharyya, Wenhao Yu, Sasha Luccioni, Paulo Villegas, Fedor Zhdanov, Tony Lee, Nadav Timor, Jennifer Ding, Claire~S Schlesinger, Hailey Schoelkopf, Jan Ebert, Tri Dao, Mayank Mishra, Alex Gu, Carolyn~Jane Anderson, Brendan Dolan-Gavitt, Danish Contractor, Siva Reddy, Daniel Fried, Dzmitry Bahdanau, Yacine Jernite, Carlos~Mu{\~n}oz Ferrandis, Sean Hughes, Thomas Wolf, Arjun Guha, Leandro~Von Werra, and Harm de~Vries. 2023{\natexlab{b}}.
\newblock \href {https://openreview.net/forum?id=KoFOg41haE} {{StarCoder}: may the source be with you!}
\newblock \emph{TMLR}.

\bibitem[{Liu et~al.(2024{\natexlab{a}})Liu, Lin, Hewitt, Paranjape, Bevilacqua, Petroni, and Liang}]{Liu2023Lost}
Nelson~F. Liu, Kevin Lin, John Hewitt, Ashwin Paranjape, Michele Bevilacqua, Fabio Petroni, and Percy Liang. 2024{\natexlab{a}}.
\newblock Lost in the middle: How language models use long contexts.
\newblock \emph{TACL}, 12:157–--173.

\bibitem[{Liu et~al.(2021)Liu, Chen, Xie, Siow, and Liu}]{Liu2021Retrieval}
Shangqing Liu, Yu~Chen, Xiaofei Xie, Jing~Kai Siow, and Yang Liu. 2021.
\newblock Retrieval-augmented generation for code summarization via hybrid {GNN}.
\newblock In \emph{ICLR}, Virtual. OpenReview.net.

\bibitem[{Liu et~al.(2024{\natexlab{b}})Liu, Xu, and McAuley}]{Liu2023RepoBench}
Tianyang Liu, Canwen Xu, and Julian~J. McAuley. 2024{\natexlab{b}}.
\newblock {RepoBench}: Benchmarking repository-level code auto-completion systems.
\newblock In \emph{ICLR}, Vienna, Austria. OpenReview.net.

\bibitem[{Liu et~al.(2023)Liu, Lai, Yu, Xu, Zeng, Du, Zhang, Dong, and Tang}]{Liu2023WebGLM}
Xiao Liu, Hanyu Lai, Hao Yu, Yifan Xu, Aohan Zeng, Zhengxiao Du, Peng Zhang, Yuxiao Dong, and Jie Tang. 2023.
\newblock {WebGLM}: Towards an efficient web-enhanced question answering system with human preferences.
\newblock In \emph{KDD}, pages 4549--4560, Long Beach, CA, USA. ACM.

\bibitem[{Lu et~al.(2022)Lu, Duan, Han, Guo, Hwang, and Svyatkovskiy}]{lu-etal-2022-reacc}
Shuai Lu, Nan Duan, Hojae Han, Daya Guo, Seung-won Hwang, and Alexey Svyatkovskiy. 2022.
\newblock \href {https://doi.org/10.18653/v1/2022.acl-long.431} {{R}e{ACC}: A retrieval-augmented code completion framework}.
\newblock In \emph{Proceedings of the 60th Annual Meeting of the Association for Computational Linguistics (Volume 1: Long Papers)}, pages 6227--6240, Dublin, Ireland. Association for Computational Linguistics.

\bibitem[{Lu et~al.(2021)Lu, Guo, Ren, Huang, Svyatkovskiy, Blanco, Clement, Drain, Jiang, Tang, Li, Zhou, Shou, Zhou, Tufano, Gong, Zhou, Duan, Sundaresan, Deng, Fu, and Liu}]{Lu2021CodeXGLUE}
Shuai Lu, Daya Guo, Shuo Ren, Junjie Huang, Alexey Svyatkovskiy, Ambrosio Blanco, Colin~B. Clement, Dawn Drain, Daxin Jiang, Duyu Tang, Ge~Li, Lidong Zhou, Linjun Shou, Long Zhou, Michele Tufano, Ming Gong, Ming Zhou, Nan Duan, Neel Sundaresan, Shao~Kun Deng, Shengyu Fu, and Shujie Liu. 2021.
\newblock {CodeXGLUE}: A machine learning benchmark dataset for code understanding and generation.
\newblock In \emph{NeurIPS}, pages 1--16, Virtual. Curran Associates, Inc.

\bibitem[{Mallen et~al.(2023)Mallen, Asai, Zhong, Das, Khashabi, and Hajishirzi}]{mallen-etal-2023-trust}
Alex Mallen, Akari Asai, Victor Zhong, Rajarshi Das, Daniel Khashabi, and Hannaneh Hajishirzi. 2023.
\newblock \href {https://doi.org/10.18653/v1/2023.acl-long.546} {When not to trust language models: Investigating effectiveness of parametric and non-parametric memories}.
\newblock In \emph{Proceedings of the 61st Annual Meeting of the Association for Computational Linguistics (Volume 1: Long Papers)}, pages 9802--9822, Toronto, Canada. Association for Computational Linguistics.

\bibitem[{Nakano et~al.(2021)Nakano, Hilton, Balaji, Wu, Ouyang, Kim, Hesse, Jain, Kosaraju, Saunders, Jiang, Cobbe, Eloundou, Krueger, Button, Knight, Chess, and Schulman}]{Nakano2021WebGPT}
Reiichiro Nakano, Jacob Hilton, Suchir Balaji, Jeff Wu, Long Ouyang, Christina Kim, Christopher Hesse, Shantanu Jain, Vineet Kosaraju, William Saunders, Xu~Jiang, Karl Cobbe, Tyna Eloundou, Gretchen Krueger, Kevin Button, Matthew Knight, Benjamin Chess, and John Schulman. 2021.
\newblock {WebGPT}: Browser-assisted question-answering with human feedback.
\newblock \emph{CoRR}, 2112.09332:1--32.

\bibitem[{Nijkamp et~al.(2023{\natexlab{a}})Nijkamp, Hayashi, Xiong, Savarese, and Zhou}]{Nijkamp2023CodeGen2}
Erik Nijkamp, Hiroaki Hayashi, Caiming Xiong, Silvio Savarese, and Yingbo Zhou. 2023{\natexlab{a}}.
\newblock {CodeGen2}: Lessons for training {LLMs} on programming and natural languages.
\newblock \emph{CoRR}, 2305.02309:1--12.

\bibitem[{Nijkamp et~al.(2023{\natexlab{b}})Nijkamp, Pang, Hayashi, Tu, Wang, Zhou, Savarese, and Xiong}]{Nijkamp2023CodeGen}
Erik Nijkamp, Bo~Pang, Hiroaki Hayashi, Lifu Tu, Huan Wang, Yingbo Zhou, Silvio Savarese, and Caiming Xiong. 2023{\natexlab{b}}.
\newblock {CodeGen}: An open large language model for code with multi-turn program synthesis.
\newblock In \emph{ICLR}, Kigali, Rwanda. OpenReview.net.

\bibitem[{OpenAI(2023)}]{gpt4}
OpenAI. 2023.
\newblock {GPT-4} technical report.
\newblock https://cdn.openai.com/papers/gpt-4.pdf.

\bibitem[{Ouyang et~al.(2022)Ouyang, Wu, Jiang, Almeida, Wainwright, Mishkin, Zhang, Agarwal, Slama, Ray, Schulman, Hilton, Kelton, Miller, Simens, Askell, Welinder, Christiano, Leike, and Lowe}]{Ouyang2022Training}
Long Ouyang, Jeffrey Wu, Xu~Jiang, Diogo Almeida, Carroll~L. Wainwright, Pamela Mishkin, Chong Zhang, Sandhini Agarwal, Katarina Slama, Alex Ray, John Schulman, Jacob Hilton, Fraser Kelton, Luke Miller, Maddie Simens, Amanda Askell, Peter Welinder, Paul~F. Christiano, Jan Leike, and Ryan Lowe. 2022.
\newblock Training language models to follow instructions with human feedback.
\newblock In \emph{NeurIPS}, volume~35, pages 27730--27744. Curran Associates, Inc.

\bibitem[{Parvez et~al.(2021)Parvez, Ahmad, Chakraborty, Ray, and Chang}]{parvez-etal-2021-retrieval-augmented}
Md~Rizwan Parvez, Wasi Ahmad, Saikat Chakraborty, Baishakhi Ray, and Kai-Wei Chang. 2021.
\newblock \href {https://doi.org/10.18653/v1/2021.findings-emnlp.232} {Retrieval augmented code generation and summarization}.
\newblock In \emph{Findings of the Association for Computational Linguistics: EMNLP 2021}, pages 2719--2734, Punta Cana, Dominican Republic. Association for Computational Linguistics.

\bibitem[{Pei et~al.(2023)Pei, Zhao, Lausen, Zha, and Karypis}]{Pei2023Better}
Hengzhi Pei, Jinman Zhao, Leonard Lausen, Sheng Zha, and George Karypis. 2023.
\newblock Better context makes better code language models: A case study on function call argument completion.
\newblock In \emph{AAAI}, pages 5230--5238, Washington, DC, USA. AAAI Press.

\bibitem[{Peng et~al.(2022)Peng, Gao, Li, Gao, Lo, Zhang, and Lyu}]{Peng2022Static}
Yun Peng, Cuiyun Gao, Zongjie Li, Bowei Gao, David Lo, Qirun Zhang, and Michael~R. Lyu. 2022.
\newblock Static inference meets deep learning: A hybrid type inference approach for {Python}.
\newblock In \emph{ICSE}, pages 2019--2030, Pittsburgh, PA, USA. ACM.

\bibitem[{Peng et~al.(2023)Peng, Li, Gu, Li, Wang, Gao, and Lyu}]{Peng2023Revisiting}
Yun Peng, Shuqing Li, Wenwei Gu, Yichen Li, Wenxuan Wang, Cuiyun Gao, and Michael~R. Lyu. 2023.
\newblock Revisiting, benchmarking and exploring {API} recommendation: How far are we?
\newblock \emph{IEEE Trans. Softw.}, 49(4):1876--1897.

\bibitem[{Proksch et~al.(2015)Proksch, Lerch, and Mezini}]{Proksch2015Intelligent}
Sebastian Proksch, Johannes Lerch, and Mira Mezini. 2015.
\newblock Intelligent code completion with {Bayesian} networks.
\newblock \emph{ACM Trans. Softw. Eng. Methodol.}, 25(1):3:1--3:31.

\bibitem[{Ram et~al.(2023)Ram, Levine, Dalmedigos, Muhlgay, Shashua, Leyton{-}Brown, and Shoham}]{Ram2023In}
Ori Ram, Yoav Levine, Itay Dalmedigos, Dor Muhlgay, Amnon Shashua, Kevin Leyton{-}Brown, and Yoav Shoham. 2023.
\newblock In-context retrieval-augmented language models.
\newblock \emph{TACL}, 11:1316--1331.

\bibitem[{Raychev et~al.(2016)Raychev, Bielik, and Vechev}]{Raychev2016Probabilistic}
Veselin Raychev, Pavol Bielik, and Martin~T. Vechev. 2016.
\newblock Probabilistic model for code with decision trees.
\newblock In \emph{OOPSLA}, pages 731--747, Amsterdam, The Netherlands. ACM.

\bibitem[{Raychev et~al.(2014)Raychev, Vechev, and Yahav}]{Raychev2014Code}
Veselin Raychev, Martin~T. Vechev, and Eran Yahav. 2014.
\newblock Code completion with statistical language models.
\newblock \emph{ACM SIGPLAN Notices}, 49(6):419--428.

\bibitem[{Robertson and Zaragoza(2009)}]{Robertson2009Probabilistic}
Stephen Robertson and Hugo Zaragoza. 2009.
\newblock The probabilistic relevance framework: {BM}25 and beyond.
\newblock \emph{Foundations and Trends{\textregistered} in Information Retrieval}, 3(4):333--389.

\bibitem[{Rozi{\`{e}}re et~al.(2023)Rozi{\`{e}}re, Gehring, Gloeckle, Sootla, Gat, Tan, Adi, Liu, Remez, Rapin, Kozhevnikov, Evtimov, Bitton, Bhatt, Canton{-}Ferrer, Grattafiori, Xiong, D{\'{e}}fossez, Copet, Azhar, Touvron, Martin, Usunier, Scialom, and Synnaeve}]{Baptiste2023codellama}
Baptiste Rozi{\`{e}}re, Jonas Gehring, Fabian Gloeckle, Sten Sootla, Itai Gat, Xiaoqing~Ellen Tan, Yossi Adi, Jingyu Liu, Tal Remez, J{\'{e}}r{\'{e}}my Rapin, Artyom Kozhevnikov, Ivan Evtimov, Joanna Bitton, Manish Bhatt, Cristian Canton{-}Ferrer, Aaron Grattafiori, Wenhan Xiong, Alexandre D{\'{e}}fossez, Jade Copet, Faisal Azhar, Hugo Touvron, Louis Martin, Nicolas Usunier, Thomas Scialom, and Gabriel Synnaeve. 2023.
\newblock {Code Llama}: Open foundation models for code.
\newblock \emph{CoRR}, 2308.12950:1--48.

\bibitem[{Shen et~al.(2023)Shen, Zhang, Chen, Zan, Geng, Fu, Zeng, Yu, Ji, Zhao, Guo, and Wang}]{Shen2023PanGu}
Bo~Shen, Jiaxin Zhang, Taihong Chen, Daoguang Zan, Bing Geng, An~Fu, Muhan Zeng, Ailun Yu, Jichuan Ji, Jingyang Zhao, Yuenan Guo, and Qianxiang Wang. 2023.
\newblock {PanGu-Coder2}: Boosting large language models for code with ranking feedback.
\newblock \emph{CoRR}, 2307.14936:1--15.

\bibitem[{Shi et~al.(2023)Shi, Min, Yasunaga, Seo, James, Lewis, Zettlemoyer, and Yih}]{Shi2023REPLUG}
Weijia Shi, Sewon Min, Michihiro Yasunaga, Minjoon Seo, Rich James, Mike Lewis, Luke Zettlemoyer, and Wen{-}tau Yih. 2023.
\newblock {REPLUG}: Retrieval-augmented black-box language models.
\newblock \emph{CoRR}, 2301.12652:1--12.

\bibitem[{Shrivastava et~al.(2023{\natexlab{a}})Shrivastava, Kocetkov, de~Vries, Bahdanau, and Scholak}]{Shrivastava2023RepoFusion}
Disha Shrivastava, Denis Kocetkov, Harm de~Vries, Dzmitry Bahdanau, and Torsten Scholak. 2023{\natexlab{a}}.
\newblock {RepoFusion}: Training code models to understand your repository.
\newblock \emph{CoRR}, 2306.10998:1--15.

\bibitem[{Shrivastava et~al.(2023{\natexlab{b}})Shrivastava, Larochelle, and Tarlow}]{Shrivastava2023Repository}
Disha Shrivastava, Hugo Larochelle, and Daniel Tarlow. 2023{\natexlab{b}}.
\newblock Repository-level prompt generation for large language models of code.
\newblock In \emph{ICML}, pages 31693--31715, Honolulu, HI, USA. PMLR.

\bibitem[{Sun et~al.(2023)Sun, Wang, Tay, Yang, and Zhou}]{Sun2023Recitation}
Zhiqing Sun, Xuezhi Wang, Yi~Tay, Yiming Yang, and Denny Zhou. 2023.
\newblock Recitation-augmented language models.
\newblock In \emph{ICLR}, Kigali, Rwanda. OpenReview.net.

\bibitem[{Svyatkovskiy et~al.(2019)Svyatkovskiy, Zhao, Fu, and Sundaresan}]{Svyatkovskiy2019Pythia}
Alexey Svyatkovskiy, Ying Zhao, Shengyu Fu, and Neel Sundaresan. 2019.
\newblock Pythia: {AI}-assisted code completion system.
\newblock In \emph{KDD}, pages 2727--2735, Anchorage, AK, USA. ACM.

\bibitem[{Touvron et~al.(2023)Touvron, Martin, Stone, Albert, Almahairi, Babaei, Bashlykov, Batra, Bhargava, Bhosale, Bikel, Blecher, Canton{-}Ferrer, Chen, Cucurull, Esiobu, Fernandes, Fu, Fu, Fuller, Gao, Goswami, Goyal, Hartshorn, Hosseini, Hou, Inan, Kardas, Kerkez, Khabsa, Kloumann, Korenev, Koura, Lachaux, Lavril, Lee, Liskovich, Lu, Mao, Martinet, Mihaylov, Mishra, Molybog, Nie, Poulton, Reizenstein, Rungta, Saladi, Schelten, Silva, Smith, Subramanian, Tan, Tang, Taylor, Williams, Kuan, Xu, Yan, Zarov, Zhang, Fan, Kambadur, Narang, Rodriguez, Stojnic, Edunov, and Scialom}]{Touvron2023llama2}
Hugo Touvron, Louis Martin, Kevin Stone, Peter Albert, Amjad Almahairi, Yasmine Babaei, Nikolay Bashlykov, Soumya Batra, Prajjwal Bhargava, Shruti Bhosale, Dan Bikel, Lukas Blecher, Cristian Canton{-}Ferrer, Moya Chen, Guillem Cucurull, David Esiobu, Jude Fernandes, Jeremy Fu, Wenyin Fu, Brian Fuller, Cynthia Gao, Vedanuj Goswami, Naman Goyal, Anthony Hartshorn, Saghar Hosseini, Rui Hou, Hakan Inan, Marcin Kardas, Viktor Kerkez, Madian Khabsa, Isabel Kloumann, Artem Korenev, Punit~Singh Koura, Marie{-}Anne Lachaux, Thibaut Lavril, Jenya Lee, Diana Liskovich, Yinghai Lu, Yuning Mao, Xavier Martinet, Todor Mihaylov, Pushkar Mishra, Igor Molybog, Yixin Nie, Andrew Poulton, Jeremy Reizenstein, Rashi Rungta, Kalyan Saladi, Alan Schelten, Ruan Silva, Eric~Michael Smith, Ranjan Subramanian, Xiaoqing~Ellen Tan, Binh Tang, Ross Taylor, Adina Williams, Jian~Xiang Kuan, Puxin Xu, Zheng Yan, Iliyan Zarov, Yuchen Zhang, Angela Fan, Melanie Kambadur, Sharan Narang, Aur{\'{e}}lien Rodriguez, Robert Stojnic, Sergey Edunov,
  and Thomas Scialom. 2023.
\newblock Llama 2: Open foundation and fine-tuned chat models.
\newblock \emph{CoRR}, 2307.09288:1--77.

\bibitem[{Trivedi et~al.(2023)Trivedi, Balasubramanian, Khot, and Sabharwal}]{trivedi-etal-2023-interleaving}
Harsh Trivedi, Niranjan Balasubramanian, Tushar Khot, and Ashish Sabharwal. 2023.
\newblock \href {https://doi.org/10.18653/v1/2023.acl-long.557} {Interleaving retrieval with chain-of-thought reasoning for knowledge-intensive multi-step questions}.
\newblock In \emph{Proceedings of the 61st Annual Meeting of the Association for Computational Linguistics (Volume 1: Long Papers)}, pages 10014--10037, Toronto, Canada. Association for Computational Linguistics.

\bibitem[{Tufano et~al.(2023)Tufano, Pascarella, and Bavota}]{Tufano2023Automating}
Rosalia Tufano, Luca Pascarella, and Gabriele Bavota. 2023.
\newblock Automating code-related tasks through transformers: The impact of pre-training.
\newblock In \emph{ICSE}, pages 2425--2437, Melbourne, Australia. IEEE.

\bibitem[{van Rossum and Lehtosalo(2022)}]{pep484}
{\L}ukasz Langa~Guido van Rossum and Jukka Lehtosalo. 2022.
\newblock {PEP} 484 – type hints.
\newblock https://peps.python.org/pep-0484/.

\bibitem[{Wang et~al.(2023)Wang, Hu, Gao, Jin, Xie, Huang, Lei, and Deng}]{Wang2023How}
Chaozheng Wang, Junhao Hu, Cuiyun Gao, Yu~Jin, Tao Xie, Hailiang Huang, Zhenyu Lei, and Yuetang Deng. 2023.
\newblock How practitioners expect code completion?
\newblock In \emph{FSE}, pages 1294--1306, San Francisco, CA, USA. {ACM}.

\bibitem[{Wang et~al.(2021)Wang, Wang, Joty, and Hoi}]{wang-etal-2021-codet5}
Yue Wang, Weishi Wang, Shafiq Joty, and Steven~C.H. Hoi. 2021.
\newblock \href {https://doi.org/10.18653/v1/2021.emnlp-main.685} {{C}ode{T}5: Identifier-aware unified pre-trained encoder-decoder models for code understanding and generation}.
\newblock In \emph{Proceedings of the 2021 Conference on Empirical Methods in Natural Language Processing}, pages 8696--8708, Online and Punta Cana, Dominican Republic. Association for Computational Linguistics.

\bibitem[{Yu et~al.(2024)Yu, Shen, Ran, Zhang, Zhang, Ma, Liang, Li, Wang, and Xie}]{Yu2024CoderEval}
Hao Yu, Bo~Shen, Dezhi Ran, Jiaxin Zhang, Qi~Zhang, Yuchi Ma, Guangtai Liang, Ying Li, Qianxiang Wang, and Tao Xie. 2024.
\newblock {CoderEval}: A benchmark of pragmatic code generation with generative pre-trained models.
\newblock In \emph{ICSE}, pages 37:1--37:12, Lisbon, Portugal. ACM.

\bibitem[{Yu et~al.(2023)Yu, Iter, Wang, Xu, Ju, Sanyal, Zhu, Zeng, and Jiang}]{Yu2023Generate}
Wenhao Yu, Dan Iter, Shuohang Wang, Yichong Xu, Mingxuan Ju, Soumya Sanyal, Chenguang Zhu, Michael Zeng, and Meng Jiang. 2023.
\newblock Generate rather than retrieve: Large language models are strong context generators.
\newblock In \emph{ICLR}, Kigali, Rwanda. OpenReview.net.

\bibitem[{Zhang et~al.(2023)Zhang, Chen, Zhang, Keung, Liu, Zan, Mao, Lou, and Chen}]{Zhang2023RepoCoder}
Fengji Zhang, Bei Chen, Yue Zhang, Jacky Keung, Jin Liu, Daoguang Zan, Yi~Mao, Jian{-}Guang Lou, and Weizhu Chen. 2023.
\newblock \href {https://aclanthology.org/2023.emnlp-main.151} {{RepoCoder}: Repository-level code completion through iterative retrieval and generation}.
\newblock In \emph{EMNLP}, pages 2471--2484, Singapore. Association for Computational Linguistics.

\bibitem[{Zhang et~al.(2020)Zhang, Wang, Zhang, Sun, and Liu}]{Zhang2020Retrieval}
Jian Zhang, Xu~Wang, Hongyu Zhang, Hailong Sun, and Xudong Liu. 2020.
\newblock Retrieval-based neural source code summarization.
\newblock In \emph{ICSE}, pages 1385--1397, Seoul, South Korea. ACM.

\bibitem[{Zheng et~al.(2023)Zheng, Xia, Zou, Dong, Wang, Xue, Shen, Wang, Wang, Li, Su, Yang, and Tang}]{Zheng2023CodeGeeX}
Qinkai Zheng, Xiao Xia, Xu~Zou, Yuxiao Dong, Shan Wang, Yufei Xue, Lei Shen, Zihan Wang, Andi Wang, Yang Li, Teng Su, Zhilin Yang, and Jie Tang. 2023.
\newblock {CodeGeeX}: A pre-trained model for code generation with multilingual benchmarking on {HumanEval-X}.
\newblock In \emph{KDD}, pages 5673--5684, Long Beach, CA, USA. ACM.

\bibitem[{Zhou et~al.(2023)Zhou, Yu, Fan, Huang, and Yang}]{Zhou2023Towards}
Ziyi Zhou, Huiqun Yu, Guisheng Fan, Zijie Huang, and Kang Yang. 2023.
\newblock Towards retrieval-based neural code summarization: A meta-learning approach.
\newblock \emph{IEEE Trans. Software Eng.}, 49(4):3008--3031.

\end{thebibliography}

%\clearpage
\appendix

\section{Dataflow Graph Construction}
\label{appendix:dfg}

We first parse the Python code into an AST by tree-sitter,\footnote{\url{https://github.com/tree-sitter/tree-sitter}} which is feasible to parse incomplete code snippets.
Based on the AST, we extract variables as the entity set and identify type-sensitive relations as triplets, forming our DFG.
The implementation involves specific Python syntax features and is presented in our published code.

\begin{figure}[!t]
    \centering
    \includegraphics[width=0.7\linewidth]{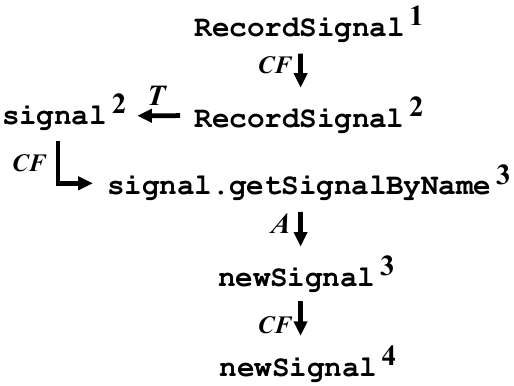}
    \caption{An example of our DFG, which corresponds to the unfinished code in Figure~\ref{fig:example}. 
    The numbers labeled in the DFG correspond to the line numbers of the variables. 
    The labels on the edges are the initials of the relation names defined in Section~\ref{sec:dfa}.
    }
    \label{fig:dfg}
\end{figure}

The type-sensitive relations are listed in Table~\ref{tab:rels}.
We also introduce an example to visualize our DFG construction, as shown in Figure~\ref{fig:dfg}.
The extracted variables include identifiers (e.g., \texttt{newSignal}) and attributes (e.g., \texttt{signal.getSignalByName}). 
The type-sensitive relations are identified as described in Section~\ref{sec:dfa}.
For example, the parameter \texttt{signal: RecordSignal} indicates a triplet (\texttt{RecordSignal}, \textit{Typeof}, \texttt{signal}).
The assignment statement forms a triplet (\texttt{signal.getSignalByName}, \textit{Assigns}, \texttt{newSignal}), and the parameter \texttt{newChannelName} is a type-insensitive relation pruned in our DFG.

\section{Generation of Background Knowledge}
\label{appendix:gbk}

Algorithms~\ref{alg:gbk} and~\ref{alg:obk} describe the generation process of the relevant background knowledge (Section~\ref{sec:generation}), where the time complexities are $\mathcal{O}(|E_o| \times (|M|+|P|))$ and $\mathcal{O}(|M|+|P|)$, respectively.

\section{Details of Experiment Setup}

\subsection{Details of Dataset Construction}
\label{appendix:recceval}

We collect the projects that are first released on PyPI between 2023-01-01 to 2023-04-28, which is after the releases of pre-training corpora \cite{husain2019codesearchnet, Chen2021Codex, Kocetkov2022Stack}.
We pick the projects with permissive licenses (i.e., MIT, Apache, and BSD) and filter out those that have fewer than 6 or more than 100 Python code files.
We identify the usages of local imported resources and randomly select a subsequent token as the cursor position.
The context before the cursor is the input, while the current line after the cursor is the reference.
For the diversity of \recceval, we limit the maximum number of examples to one per code file and 10 per repository.
Moreover, we ensure that the reference is not in the unfinished code and feed the examples to StarCoderBase-1B model \cite{Li2023Starcoder} to remove the exact matches \cite{Ding2023CrossCodeEval}, which excludes strong clues in the unfinished code to make \recceval more challenging.

\begin{table}[!ht]
\centering
{\small
\begin{tabular}{l|c|c}
    \toprule
    Relations & Examples & Triplets \\
    
    \midrule
    \textit{assigns}    & \texttt{v = u} & (\texttt{u}, \textit{assigns}, \texttt{v}) \\
    \textit{as}         & \texttt{with f() as v} & (\texttt{f}, \textit{as}, \texttt{v}) \\
    \textit{refers}     & \texttt{u.v} & (\texttt{u}, \textit{refers}, \texttt{u.v}) \\
    \textit{typeof}     & \texttt{def f() -> v} & (\texttt{v}, \textit{typeof}, \texttt{f}) \\
    \textit{inherits}   & \texttt{class v(u)} & (\texttt{u}, \textit{inherits}, \texttt{v}) \\
    
    \bottomrule
\end{tabular}
}
\caption{Illustrations of type-sensitive relations.}
\label{tab:rels}
\end{table}

\subsection{Implementation Details of Baselines}
\label{appendix:baseline}

We describe more implementation details of CCFinder, RG-1, and RepoCoder, which are in line with the experiment setup in their papers:
\begin{compactitem}
\item\textbf{CCFinder.}
Because CCFinder is not open source, we reproduce it according to its paper.
We do not limit the number of retrieved code entities, as the cross-file context would be truncated if it exceeds the maximum length.
We also re-order the retrieved entities, ensuring the entities from the same source file follow the original code order.

\item\textbf{RG-1 and RepoCoder.}
In our experiments, we use a sparse bag-of-words model as their retriever, which calculates text similarity using the Jaccard index and achieves equivalent performance to the dense retriever.
The line length of the sliding window and the sliding size are set to 20 and 10, respectively.
According to the maximum input length of code LMs, the maximum number of the retrieved code snippets in prompts is set to 40 for the StarCoder model and 10 for other models.
The number of iterations of RepoCoder is set to 2.
\end{compactitem}

\begin{algorithm}[!t]
    \caption{$Generate\_BK(E_r, E_o, n)$}
    \label{alg:gbk}

    \KwIn{the relevant import entities $E_r$, other import entities $E_o$, and the number of allocated tokens $n$}
    \KwOut{background knowledge $bk$}

    \tcp{\small the primary background knowledge}
    $E_c \leftarrow E_r$\;
    $bk \leftarrow Organize\_BK(E_c)$\;
    \ForEach{$e \in E_o$}
    {
        Append $e$ to $E_c$\;
        $temp \leftarrow Organize\_BK(E_c)$\;
        \If{$Length(temp) \leq n$}
        {
            \tcp{\small as many entities as possible}
            $bk \leftarrow temp$\;
        }
        \Else
        {
            \tcp{\small prevents the primary background knowledge from being truncated}
            \textbf{break}\;
        }
    }

    \If{$Length(bk) > n$}
    {
        $bk \leftarrow$ the first $n$ tokens of $bk$\;
    }
    \Return $bk$

\end{algorithm}

\begin{algorithm}[!t]
    \caption{$Organize\_BK(E_c)$}
    \label{alg:obk}
    
    \KwIn{the ordered import entities $E_c$}
    \KwOut{background knowledge $bk$}

    \tcp{\small let $G_m = (M, P)$, where $M$ is the vertex set, and $P$ is the edge set}
    $G_m \leftarrow$ retrieve all dependent entities of $E_c$ and group them in modules\;
    $M_e \leftarrow$ the corresponding modules of $E_c$\;

    \tcp{\small set the priorities of modules $M$}
    Initialize an empty dictionary $priorities$\;
    $rank \leftarrow 1$\;
    \ForEach{$m \in M_e$}
    {
        \tcp{\small priorities \romannumeral2}
        $priorities[m] \leftarrow rank$\;
        $rank \leftarrow rank + 1$\;
    }

    $M_d \leftarrow M - M_e$\;
    \tcp{\small details omitted: pass priorities from $M_e$ to the dependent modules $M_d$}
    $\forall m \in M_d$, $priorities[m] \leftarrow$ the minimum priority of the direct predecessors of $m$;

    \tcp{\small priority \romannumeral1: pseudo-topological sort}
    Initialize an empty list $M_s$\;
    \While{$M$ is not empty}
    {
        $M_c \leftarrow$ the candidate modules with the least direct predecessors in $G_m$\;
        \tcp{\small alphabetical order for equality}
        $m \leftarrow$ the module $m \in M_c$ with the maximum $priorities[m]$\;
        Pop $m$ from $M$ and append it to $M_s$\;
    }

    $bk \leftarrow$ combine all modules in the reverse order of $M_s$\;
    \Return $bk$
    
\end{algorithm}

\begin{table*}
\centering
\resizebox{\linewidth}{!}
{
\begin{tabular}{l|cccc|cccc|cccc}
    \toprule
    \multirow{3}{*}{Methods} & \multicolumn{4}{c|}{CodeGen-2.7B} & \multicolumn{4}{c|}{CodeGen-6.1B} & \multicolumn{4}{c}{CodeGen-16.1B} \\
    \cmidrule(lr){2-5} \cmidrule(lr){6-9} \cmidrule(lr){10-13}
    & EM & ES & ID.EM & F1 & EM & ES & ID.EM & F1 & EM & ES & ID.EM & F1 \\

    \midrule
    Zero-Shot   & \ \ 5.44 & 57.85 & 11.71 & 42.22 & \ \ 6.57 & 59.01 & 13.13 & 44.11 & \ \ 7.05 & 59.88 & 13.88 & 45.27 \\
    CCFinder-1  & 14.30 & 63.18 & 22.51 & 51.28 & 16.21 & 65.00 & 24.58 & 53.70 & 17.19 & 65.57 & 26.19 & 55.36 \\
    CCFinder-2  & 11.41 & 61.74 & 19.47 & 48.92 & 13.21 & 63.23 & 21.39 & 51.17 & 14.15 & 63.89 & 22.59 & 52.17 \\
    RG-1        & 12.68 & 63.87 & 21.58 & 51.89 & 14.82 & 65.12 & 23.53 & 53.54 & 15.27 & 65.87 & 24.65 & 54.76 \\
    RepoCoder   & 14.07 & 65.12 & 23.90 & 53.33 & 15.87 & 66.74 & 26.15 & 55.80 & 17.04 & 67.69 & 27.62 & 57.36 \\
    
    \methodname & \textbf{18.99} & \textbf{65.52} & \textbf{27.50} & \textbf{55.07} & \textbf{22.36} & \textbf{68.06} & \textbf{31.37} & \textbf{58.60} & \textbf{22.78} & \textbf{68.09} & \textbf{32.08} & \textbf{59.40} \\
    % improvement & & & & & & & & & & & & \\
    \bottomrule
\end{tabular}
}
\caption{Performance comparison on the \crosscodeeval dataset using other CodeGen models (cf. Table~\ref{tab:exp-cceval}).}
\label{tab:exp-codegen-cceval}
\end{table*}

\begin{table*}
\centering
\resizebox{\linewidth}{!}
{
\begin{tabular}{l|cccc|cccc|cccc}
    \toprule
    \multirow{3}{*}{Methods} & \multicolumn{4}{c|}{CodeGen-2.7B} & \multicolumn{4}{c|}{CodeGen-6.1B} & \multicolumn{4}{c}{CodeGen-16.1B} \\
    \cmidrule(lr){2-5} \cmidrule(lr){6-9} \cmidrule(lr){10-13}
    & EM & ES & ID.EM & F1 & EM & ES & ID.EM & F1 & EM & ES & ID.EM & F1 \\

    \midrule
    Zero-Shot   & \ \ 6.73 & 53.30 & 13.05 & 30.65 & \ \ 8.34 & 54.77 & 14.64 & 32.60 & 10.12 & 55.84 & 16.50 & 34.17 \\
    CCFinder-1  & 20.38 & 60.80 & 28.12 & 44.83 & 23.56 & 63.07 & 31.56 & 47.90 & 24.64 & 64.17 & 32.66 & 49.28 \\
    CCFinder-2  & 17.21 & 59.13 & 24.32 & 41.58 & 19.66 & 60.77 & 26.93 & 43.73 & 20.83 & 61.85 & 28.25 & 45.11 \\
    RG-1        & 24.49 & 63.12 & 31.34 & 46.51 & 25.86 & 64.75 & 32.66 & 48.37 & 27.97 & 66.18 & 35.07 & 50.37 \\
    RepoCoder   & 27.84 & 65.07 & 35.13 & 49.71 & 29.45 & 66.62 & 36.71 & 51.67 & 31.73 & 67.94 & 38.96 & 53.64 \\
    
    \methodname & \textbf{29.42} & \textbf{65.91} & \textbf{37.63} & \textbf{53.69} & \textbf{32.05} & \textbf{67.93} & \textbf{40.83} & \textbf{56.80} & \textbf{33.76} & \textbf{69.20} & \textbf{42.38} & \textbf{58.38} \\
    % improvement & & & & & & & & & & & & \\
    \bottomrule
\end{tabular}
}
\caption{Performance comparison on the \recceval dataset using other CodeGen models (cf. Table~\ref{tab:exp-recceval}).}
\label{tab:exp-codegen-recceval}
\end{table*}

\subsection{Details of Used LMs}
\label{appendix:lms}

We categorize the used LMs into specialized code LMs, adapted code LMs, and GPT models.
The details are listed as follows:
\begin{compactitem}
\item\textbf{CodeGen} \cite{Nijkamp2023CodeGen2,Nijkamp2023CodeGen} is a family of auto-regressive LMs for program synthesis.
We use the CodeGen2.5 model with 7B parameters and the CodeGen models with 350M, 2.7B, 6.1B, and 16.1B parameters, which support a maximum context length of 2,048 (2K) tokens.
We use their mono versions, which are further trained on additional Python tokens.

\item\textbf{SantaCoder} \cite{Allal2023Santacoder} is a code model with 1.1B parameters, which supports a maximum context length of 2K tokens.

\item\textbf{StarCoder} \cite{Li2023Starcoder} is a 15.5B model trained on 80+ programming languages and further trained on Python, which supports a maximum context length of 8K tokens.

\item\textbf{Code Llama} \cite{Baptiste2023codellama} is created by further training Llama 2 \cite{Touvron2023llama2} on its code-specific datasets, which supports a maximum context length of 16k tokens.
Considering GPU member and efficiency, we chose the 7B Python specialist version named CodeLlama-7b-Python-hf and limit the maximum context length to 8K tokens.

\item\textbf{GPT models} \cite{Ouyang2022Training, gpt4} are commercial black box models released by OpenAI. 
The used API of GPT-3.5 is gpt-3.5-turbo-0613 with a maximum context length of 4K tokens.
The used API of GPT-4 is gpt-4-0613 with a maximum context length of 8K tokens.
\end{compactitem}

The analysis experiments with Code LLama and GPT models may suffer from data leakage issues.
As mentioned in Appendix~\ref{appendix:recceval}, both \crosscodeeval and our \recceval devote effort to collecting projects (e.g., released between 2023-01-01 to 2023-04-28 for \recceval) that are not in pre-training corpora. 
However, Code Llama is trained on unknown datasets between January 2023 and July 2023, and the training data of GPT models is unknown and updated.

We set the temperature of these LMs as 0 to obtain deterministic results.
The maximum generation length is set to 48 tokens, which is long enough to accomplish line completions.
An exception is RG-1, which asks LMs to generate 100 tokens since RepoCoder requires sufficient content for further retrieval.
We run StarCoder-15.5B, CodeGen-16.1B, and Code Llama-7B on an NVIDIA A800 with 80GB memory and run other LMs on an NVIDIA GeForce RTX 4090 with 24GB memory.

The instruction of GPT models is ``You are a Python expert. Please complete the last line of the following Python code:''. 
For RG-1, the instruction is ``You are a Python expert. Please complete the following Python code:''. 
We set line feed as the stop token of LM generation, except for RG-1 (still 100 tokens). 

\subsection{Details of Evaluation Metrics}
\label{appendix:metric}

\begin{compactitem}
\item\textbf{Code match.}
Given a prediction $y$ and the reference $y^*$, we assess $y$ using the exact match accuracy (EM) and the Levenshtein edit similarity (ES) \cite{Lu2021CodeXGLUE, Zhang2023RepoCoder}.
EM is calculated by an indicator function whose value is $1$ if $y = y^*$; otherwise, it is $0$.
$\text{ES} = 1 - \frac{\text{Lev}(y, y^*)}{\max(||y||, ||y^*||)}$, where $||\cdot||$ calculates the string length and $\text{Lev}()$ calculates the Levenshtein distance.

\item\textbf{Identifier match.}
Identifier exact match (ID.EM) and F1-score test the model’s ability to predict the correct APIs \cite{Ding2023CrossCodeEval}.
We parse the code to extract the identifiers from $y$ and $y^*$ and get two ordered identifier lists.
ID.EM is calculated by an indicator function whose value is $1$ if their elements are the same and in the same order; otherwise, it is $0$. Then, we transform the lists into two sets (without repeated elements and unordered), which can be used to calculate the precision, recall, and F1-score, where F1-score is a combination of precision and recall.

\item\textbf{Prompt generation time.}
To evaluate the efficiency of code completion, we record the prompt generation time, which contains the time to retrieve relevant context and the time to assemble final prompts.
Note that we ignore the time spent by code LMs in generating predictions, which is determined by the used LMs rather than the methods.
\end{compactitem}

\begin{table}[!t]
\centering
\resizebox{\columnwidth}{!}
{
    \begin{tabular}{l|c|cc}
         \toprule
         Methods & Models & \crosscodeeval & \recceval \\
         \midrule
         CCFinder & All & 74 & 68 \\      
         \multirow{4}{*}{\makecell{RG-1 \& \\RepoCoder}} & CodeGen & 227 & 217 \\
         & SantaCoder & 245 & 221 \\
         & CodeGen25 & 349 & 339 \\
         & StarCoder & 211 & 186 \\
         \methodname & All & \textbf{54} & \textbf{61} \\
         \bottomrule
    \end{tabular}
}
\caption{Preprocessing time (in milliseconds) for the repositories in \crosscodeeval and \recceval.}
\label{tab:exp-preprocess}
\end{table}

\begin{table*}
\centering
{\small
\begin{tabular}{l|cc|cc|cc}
    \toprule
    \multirow{3}{*}{Methods} & \multicolumn{2}{c|}{SantaCoder-1.1B} & \multicolumn{2}{c|}{CodeGen25-7B} & \multicolumn{2}{c}{StarCoder-15.5B} \\
    \cmidrule(lr){2-3} \cmidrule(lr){4-5} \cmidrule(lr){6-7}
    & \crosscodeeval & \recceval & \crosscodeeval & \recceval & \crosscodeeval & \recceval \\

    \midrule
    CCFinder-1  & 30 & 52 & 23 & 34 & 26 & 45 \\
    CCFinder-2  & 47 & 72 & 37 & 51 & 42 & 66 \\
    RG-1        & \textbf{15} & \textbf{15} & \textbf{18} & \textbf{20} & \textbf{12} & \textbf{15} \\
    RepoCoder   & 3,075 & 3,184 & 5,249 & 4,772 & 4,746 & 4,675 \\
    
    \methodname & 35 & 42 & 32 & 36 & 60 & 77 \\
    \bottomrule
\end{tabular}
}
\caption{Prompt generation time (in milliseconds) of each method using SantaCoder, CodeGen25, and StarCoder models (cf. Table~\ref{tab:time}).}
\label{tab:time-others}
\end{table*}

\begin{table*}
\centering
{\small
\begin{tabular}{l|cc|cc|cc}
    \toprule
    \multirow{3}{*}{Methods} & \multicolumn{2}{c|}{CodeGen-2.7B} & \multicolumn{2}{c|}{CodeGen-6.1B} & \multicolumn{2}{c}{CodeGen-16.1B} \\
    \cmidrule(lr){2-3} \cmidrule(lr){4-5} \cmidrule(lr){6-7}
    & \crosscodeeval & \recceval & \crosscodeeval & \recceval & \crosscodeeval & \recceval \\

    \midrule
    CCFinder-1  & 32 & 49 & 32 & 49 & 32 & 49 \\
    CCFinder-2  & 52 & 82 & 52 & 82 & 52 & 82 \\
    RG-1        & \textbf{14} & \textbf{15} & \textbf{19} & \textbf{14} & \textbf{12} & \textbf{13} \\
    RepoCoder   & 6,933 & 5,779 & 7,543 & 6,236 & 7,289 & 7,137 \\
    
    \methodname & 40 & 44 & 40 & 44 & 40 & 44 \\
    \bottomrule
\end{tabular}
}
\caption{Prompt generation time (in milliseconds) of each method using other CodeGen models  (cf. Table~\ref{tab:time}).}
\label{tab:time-codegen}
\end{table*}

\begin{table*}
\centering
\resizebox{\linewidth}{!}
{
\begin{tabular}{ll|cccc|cccc|cccc|cccc}
    \toprule
    \multicolumn{2}{c|}{\multirow{3}{*}{Settings}} & \multicolumn{4}{c|}{CodeGen-350M} & \multicolumn{4}{c|}{SantaCoder-1.1B} & \multicolumn{4}{c|}{CodeGen25-7B} & \multicolumn{4}{c}{StarCoder-15.5B} \\
    \cmidrule(lr){3-6} \cmidrule(lr){7-10} \cmidrule(lr){11-14} \cmidrule(lr){15-18}
    & & EM & ES & ID.EM & F1 & EM & ES & ID.EM & F1 & EM & ES & ID.EM & F1 & EM & ES & ID.EM & F1 \\

    \midrule
    \multirow{2}{*}{\crosscodeeval} & \methodname & \textbf{13.02} & \textbf{61.30} & \textbf{20.53} & \textbf{49.04} & \textbf{20.64} & \textbf{67.04} & \textbf{29.83} & \textbf{57.37} & \textbf{24.99} & 70.10 & 34.63 & 61.14 & \textbf{34.67} & \textbf{75.83} & \textbf{45.63} & \textbf{69.93} \\
    & Variant  & 12.87 & 61.20 & 20.45 & 48.76 & 20.45 & 67.03 & \textbf{29.83} & 57.36 & 24.77 & \textbf{70.22} & \textbf{34.71} & \textbf{61.37} & 33.55 & 75.58 & 44.50 & 69.33 \\

    \midrule
        
    \multirow{2}{*}{\recceval} & \methodname & \textbf{22.12} & \textbf{60.41} & \textbf{29.73} & \textbf{46.09} & \textbf{30.26} & \textbf{66.90} & \textbf{39.08} & \textbf{55.43} & \textbf{36.46} & \textbf{70.76} & \textbf{44.67} & \textbf{60.40} & \textbf{46.49} & \textbf{76.80} & \textbf{55.98} & \textbf{70.32} \\
    & Variant  & 21.34 & 60.01 & 29.05 & 45.57 & 29.36 & 66.36 & 38.04 & 54.67 & 35.51 & 70.18 & 43.82 & 59.52 & 44.47 & 75.67 & 53.92 & 68.89 \\
    \bottomrule
\end{tabular}
}
\caption{Performance comparison between \methodname and its variant (containing only the relevant \texttt{import} statements).}
\label{tab:exp-import}
\end{table*}

\begin{table}
\centering
\resizebox{\linewidth}{!}
{
\begin{tabular}{l|cccc|cccc}
    \toprule
    \multirow{3}{*}{Methods} & \multicolumn{4}{c|}{SantaCoder-1.1B} & \multicolumn{4}{c}{CodeGen25-7B} \\
    \cmidrule(lr){2-5} \cmidrule(lr){6-9} 
    & EM & ES & ID.EM & F1 & EM & ES & ID.EM & F1 \\
    
    \midrule
    Zero-Shot           & 0.53 & 57.77 & 1.84 & 46.17 & 1.35 & 59.60 & 3.08 & 49.06 \\
    CCFinder-1          & 1.84 & 60.08 & 3.83 & 51.45 & 3.08 & 62.17 & 6.00 & 54.00 \\
    RG-1                & 2.59 & 61.49 & 5.03 & 52.11 & 3.79 & 63.67 & 6.94 & 55.45 \\
    RepoCoder           & 2.81 & \textbf{62.23} & 5.59 & \textbf{53.56} & \textbf{4.32} & \textbf{64.69} & \textbf{7.58} & \textbf{56.88} \\
    
    \methodname         & \textbf{3.23} & 61.26 & \textbf{5.85} & \textbf{53.56} & 4.13 & 63.62 & 7.54 & 56.47 \\

    \quad w/o dataflow  & 2.40 & 60.23 & 4.69 & 51.40 & 3.41 & 62.46 & 6.15 & 54.19 \\
    \quad w/ steps      & 0.34 & 48.29 & 1.28 & 36.05 & 3.94 & 62.37 & 7.02 & 55.70 \\
    
    \bottomrule
\end{tabular}}
\caption{Performance comparison on the "multi-line" variant of the CrossCodeEval dataset.}
\label{tab:exp-multi}
\end{table}

\section{Additional Evaluation}

\subsection{More Performance Comparison Results}
\label{appendix:performance}

Beyond the experimental results of the main paper, we show additional evaluation results of other CodeGen models in Tables~\ref{tab:exp-codegen-cceval} and~\ref{tab:exp-codegen-recceval}.
The additional results show consistent conclusions on performance comparisons in the main paper.
Under the same architecture of the CodeGen-* models, the performance of all methods improves as the model parameters increase.
Moreover, the improvement of \methodname for zero-shot code LMs increases as the model's capability grows.
It indicates that stronger LMs can better utilize the relevant background knowledge retrieved by \methodname.

\subsection{More Efficiency Evaluation Results}
\label{appendix:time}

We also record the time spent on indexing the repositories of \crosscodeeval and \recceval, as shown in Table~\ref{tab:exp-preprocess}.
It is an offline preprocessing in RAG, which indicates the time required to activate a method.
CCFinder and \methodname build retrieval databases by statically parsing code files, which are independent of the used code LMs.
RG-1 and RepoCoder need to tokenize the code snippets within a sliding window, which requires the tokenizers of used LMs.
Note that the tokenizers of CodeGen-* models are the same.
\methodname is 3--6 times faster than RepoCoder in preprocessing time.
As the size of the repository increases, the preprocessing time grows linearly.
Therefore, RG-1 and RepoCoder may suffer from scalability challenges.

The prompt generation time of each method using other code LMs is shown in Tables~\ref{tab:time-others} and~\ref{tab:time-codegen}, which show consistent conclusions with the main paper.
For the methods with one retrieval, only the tokenizers have a subtle effect on efficiency when different models are employed.
As a result, the prompt generation time using different CodeGen-* models is the same for CCFinder, RG-1, as well as \methodname.
RepoCoder relies on RG-1 to generate sufficient content for the second retrieval, where the efficiency mainly depends on the generation time of code LMs.
In general, the generation efficiency of RepoCoder decreases as the model parameters increase.
Its average prompt generation time is more than 3 seconds on the most efficient SantaCoder model, which far exceeds the time spent by other retrieval-augmented methods.
Note that the architectures of code LMs also matter in efficiency, e.g., SantaCoder-1.1B is faster than CodeGen-350M.
The A800 GPU used to run the StarCoder-15.5B and CodeGen-16.1B models is superior to the RTX 4090 GPU used for the other models, so these are not head-to-head comparisons for RepoCoder.

\subsection{Effect of Other Import Statements}
\label{appendix:import}
As described in Section~\ref{sec:generation}, \methodname includes as many other \texttt{import} statements that are not directly relevant as possible.
To analyze the effect of other \texttt{import} statements, we implement a variant of \methodname for comparison, which adds the entities from other local \texttt{import} statements only when the primary prompt is empty. 
The experimental results are shown in Table~\ref{tab:exp-import}. 
\methodname consistently outperforms this variant, especially on StarCoder-15.5B, which has capability to understand longer code context.

\begin{figure}[!ht]
    \centering
    \includegraphics[width=\linewidth]{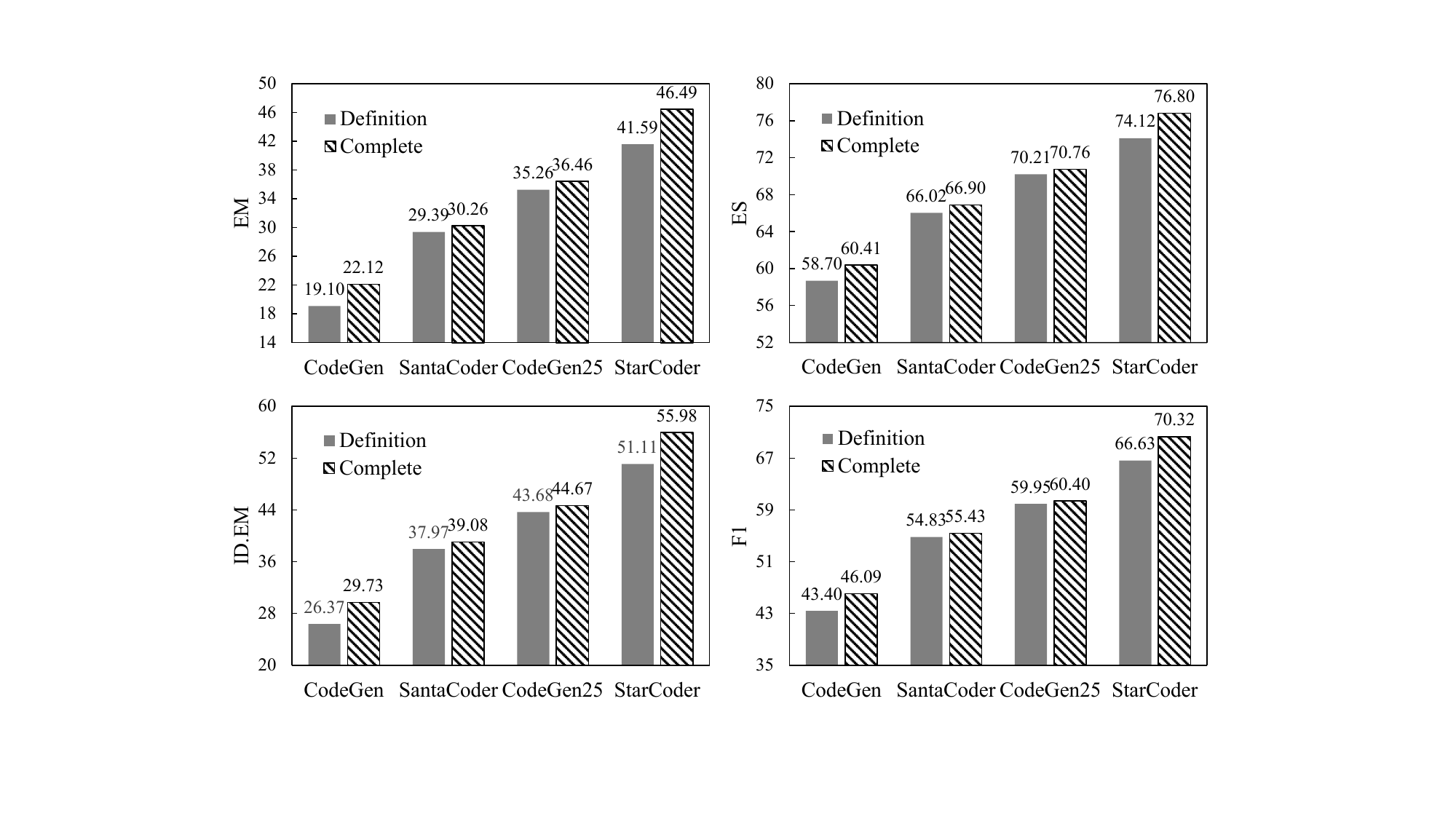}
    \caption{Performance comparison of two prompt scopes on the \recceval dataset.}
    \label{fig:scope-recc}
\end{figure}

\subsection{Evaluation of Multi-Line Completion}
\label{appendix:multi}

Although \methodname is designed for the code completion of currently edited line, it can help with multi-line completion to some extent.

We begin by discussing the benchmarks with multi-line completion targets in code repositories.
The input code of the function completion dataset in RepoEval \cite{Zhang2023RepoCoder} is truncated for text similarity-based retrieval, which cannot be parsed to get import information.
Other benchmarks \cite{Yu2024CoderEval, li2024evocodebench} are constructed for function-level code generation, which aims to generate code based on natural language descriptions, as distinct from code completion.
Since there is no available dataset for multi-line completion, we create a "multi-line" variant of the CrossCodeEval dataset \cite{Ding2023CrossCodeEval} by expanding the references to three non-empty lines.
In addition to the baselines, we evaluate two variants of \methodname including ``w/o dataflow'' as described in Section~\ref{sec:abl} and ``w/ steps'', which completes the unfinished code line by line (i.e., three times ``retrieval-then-generation").
The maximum generation length of LMs is increased to 100 tokens.

The experimental results with SantaCoder and CodeGen25 models shown in Table~\ref{tab:exp-multi} reveal that: 
(\romannumeral1) \methodname still delivers a significant improvement to the zero-shot setting, and dataflow analysis plays a positive role. 
As discussed in the Limitations section, unclear code intent in multi-line completion would hurt the accuracy and weaken the guidance of dataflow analysis.
(\romannumeral2) RepoCoder iteratively retrieves similar code snippets in the repository to narrow the gap with the intended completion target, which is slightly better than \methodname but still struggles with multi-line code completion.
(\romannumeral3) Completing line by line is not as effective as completing it all at once. 
The intermediate generated line may not point to cross-file imports in the dataflow, which would weaken the relevance of the retrieved content. 
In addition, progressive completion may lead to error accumulation.
Note that the abnormal results of ``w/ step'' with SantaCoder are attributed to the model itself, which usually generates a comment or another function definition when completing a new line.

Recent empirical studies have shown that the completion of currently edited statement is more commonly used and more acceptable \cite{Barke2023Grounded, Wang2023How}. 
In contrast, long and multi-line suggestions are usually at best dismissed out of hand and at worst distract the programmer away from their flow. 
Based on this practical evidence, we believe that our focus on single-line code completion is meaningful and sufficient. Future work could explore taking into account the dynamic changes of dataflow during the decoding process, e.g., automatically determining when it is necessary to re-retrieve to update the prompt.

\begin{figure*}[!t]
	\centering
	\subfloat[CCFinder-*.]{\includegraphics[width=.45\linewidth]{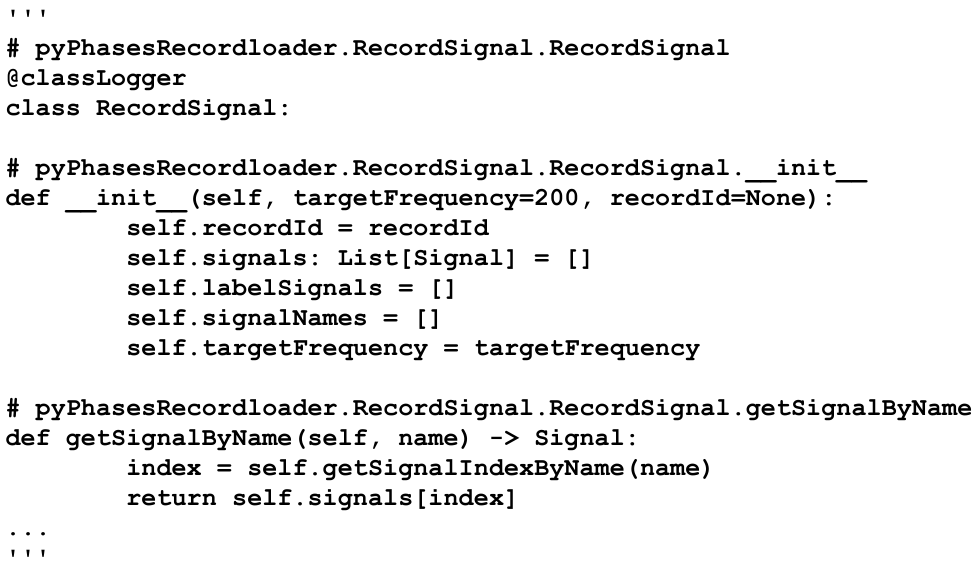}}
	\quad 
        \subfloat[RepoCoder, same as RG-1.]{\includegraphics[width=.45\linewidth]{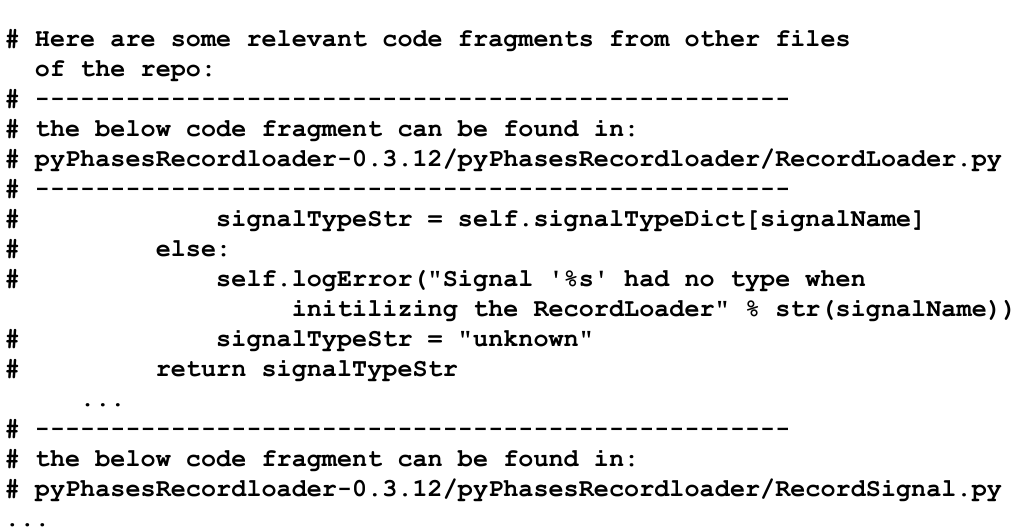}}\\
	\subfloat[Our \methodname.]{\includegraphics[width=.45\linewidth]{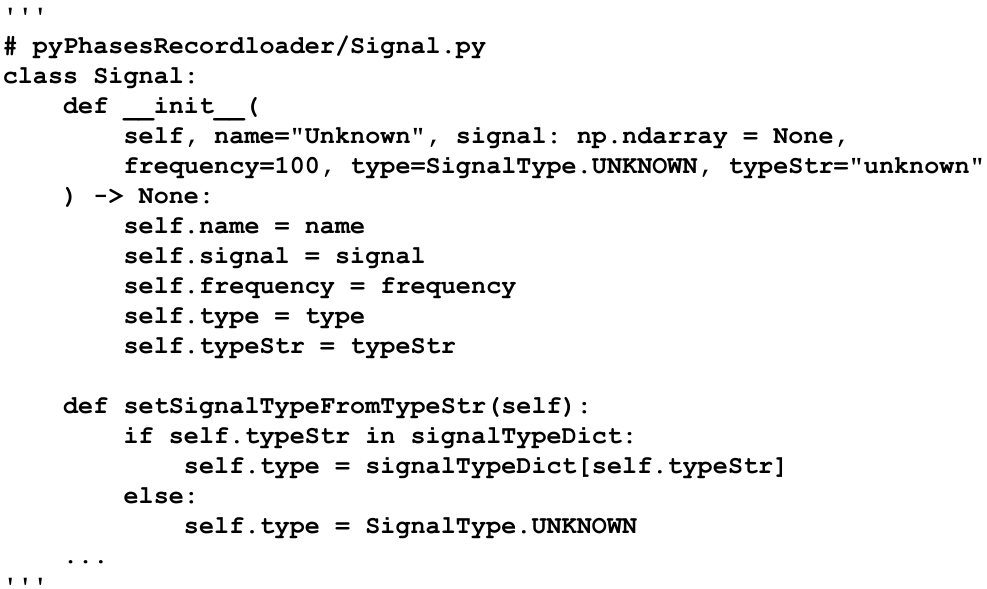}}
	\quad 
        \subfloat[\methodname with the \textit{definition} scope.]{\includegraphics[width=.45\linewidth]{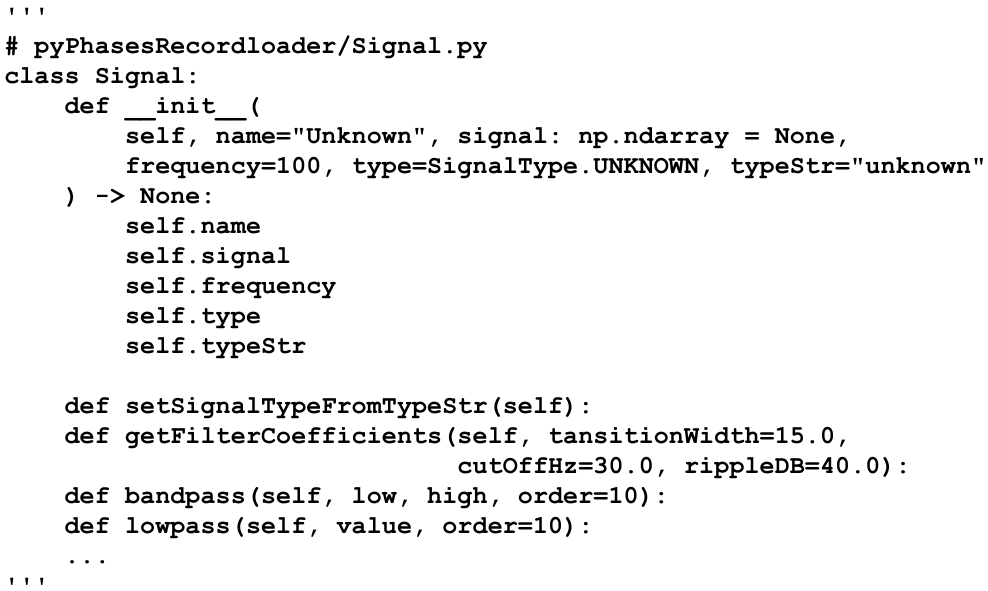}}
	\caption{Excerpts of example prompts generated by different methods.}
    \label{fig:excerpt}
\end{figure*}

\begin{table}[!t]
\centering
\resizebox{\columnwidth}{!}
{
\begin{tabular}{l|l|r}
    \toprule
    Methods & Predictions & ES \\
    \midrule
    Zero-Shot     & channel = newChannelName & 24 \\
    CCFinder-1    & type = Signal.getType(channelType) & 53 \\
    CCFinder-2    & type = Signal.getType(channelType) & 53 \\
    RG-1          & type = channelType & 36 \\
    RepoCoder     & signal = newSignal.signal.astype(channelType) & 45 \\
    \methodname   & setSignalTypeFromTypeStr() & 100 \\
    \midrule
    Ground truth  & setSignalTypeFromTypeStr() & - \\
    \bottomrule
\end{tabular}}
\caption{The example prediction of each method using the CodeGen25-7B model.}
\label{tab:prediction}
\end{table}

\section{Case Study}
\subsection{Prompt Examples}
\label{appendix:prompt}

We show the prompts generated by each method for the example unfinished code (see Figure~\ref{fig:example}).
The prompts are excerpted for viewing the individual format, as shown in Figure~\ref{fig:excerpt}.
It can be observed that the prompts generated by \methodname look like natural code, which is in line with the training corpora of code LMs.
The prediction result of each method using the CodeGen25-7B model is shown in Table~\ref{tab:prediction}, and only our \methodname generates the correct code line.

\subsection{Study on Failed Cases}
\label{appendix:fail}

Despite the superior performance achieved by \methodname, there are still many failed cases.
Based on our observations, they are caused by three major reasons: 
(\romannumeral1) The definitions of completion targets may be truncated. 
Even though the correct definition is retrieved through dataflow analysis, the completion target may be truncated due to the limited input length, e.g., target member function in a long class definition. 
(\romannumeral2) The used code LMs may be not powerful enough, which has already been demonstrated in Section~\ref{sec:adlm}.
It is crucial for code LMs to capture data dependency relations in the provided background knowledge, where the prompts are usually long Python code.
(\romannumeral3) Unclear code intent may lead to wrong generation, which is a common sore point of code completion.
For example, completing the first line of a new function is uncertain even to a programmer.

In Listings~\ref{ls-uc},~\ref{ls-tbk}, and~\ref{ls-cbk}, we show an example of reasons (\romannumeral1) and (\romannumeral3) from \crosscodeeval.
The ground truth is \texttt{gen\_begin\_reuse(input\_ids)}, while the prediction of \methodname with CodeGen25 is \texttt{in\_beam\_search = False}.
Although the member function \texttt{gen\_begin\_reuse} is in the retrieved background knowledge, it is truncated to be invisible to LMs.
Moreover, the comment \texttt{\# Start generation} is not clear enough for completion, where both \texttt{gen\_begin} and \texttt{gen\_begin\_reuse} look like rational choices.

\begin{lstlisting}[language=Python, caption=The unfinished code to be completed., label=ls-uc]
from model import ExLlama, ExLlamaCache, ExLlamaConfig
from lora import ExLlamaLora
from tokenizer import ExLlamaTokenizer
from generator import ExLlamaGenerator
import model_init

generator: ExLlamaGenerator

def cached_tokenize(text: str):
    ...

def begin_stream(...):
    global model, cache, config, generator, tokenizer

    # Settings
    max_stop_string = 2
    for ss in stop_strings:
        ...
    generator.settings = gen_settings

    # Start generation
    generator.
\end{lstlisting}

\begin{lstlisting}[language=Python, caption=The truncated background knowledge retrieved by \methodname., label=ls-tbk]
'''
# generator.py
class ExLlamaGenerator:
    class Settings:
        temperature = 0.95
        top_k = 40
        top_p = 0.65
        min_p = 0.0
        typical = 0.0
        token_repetition_penalty_max = 1.15
        token_repetition_penalty_sustain = 256
        token_repetition_penalty_decay = 128
        beams = 1
        beam_length = 1
    sequence: torch.Tensor or None
    sequence_actual: torch.Tensor or None
    settings: Settings
    beams: int or None
    max_beam_length: int
    in_beam_search: True
    disallowed_tokens: list[int] or None
    lora: ExLlamaLora or None
    def __init__(self, model, tokenizer, cache):
        self.model = model
        self.tokenizer = tokenizer
        self.cache = cache
        self.reset()
        self.model = model
        self.tokenizer = tokenizer
        self.cache = cache
    def reset(self):
        ...
    def make_rep_mask(self, penalty_max, sustain, decay):
        ...
    def batched_sample(self, logits, temperature, top_k, top_p, min_p, typical, num = 1):
        ...
    def sample_current(self, logits, num = 1):
        ...
    def sample(self, logits, temperature, top_k, top_p, min_p, typical, num = 1):
        ...
'''
\end{lstlisting}

\begin{lstlisting}[language=Python, caption=The complete background knowledge retrieved by \methodname., label=ls-cbk]
'''
# generator.py
class ExLlamaGenerator:
    class Settings:
        ...
    sequence: torch.Tensor or None
    sequence_actual: torch.Tensor or None
    settings: Settings
    beams: int or None
    max_beam_length: int
    in_beam_search: True
    disallowed_tokens: list[int] or None
    lora: ExLlamaLora or None
    def __init__(self, model, tokenizer, cache):
        ...
    def reset(self):
        ...
    def make_rep_mask(self, penalty_max, sustain, decay):
        ...
    def batched_sample(self, logits, temperature, top_k, top_p, min_p, typical, num = 1):
        ...
    def sample_current(self, logits, num = 1):
        ...
    def sample(self, logits, temperature, top_k, top_p, min_p, typical, num = 1):
        ...
    def disallow_tokens(self, tokens):
        ...
    def gen_begin(self, in_tokens, mask = None):
        ...
    def gen_begin_empty(self):
        ...
    def gen_begin_reuse(self, in_tokens, mask = None):
        self.end_beam_search()
        ...
    def gen_feed_tokens(self, in_tokens, mask = None):
        ...
    def gen_accept_token(self, token):
        ...
    def gen_rewind(self, num_tokens):
        ...
    def gen_prune_right(self, tokens, mask = None):
        ...
    def gen_prune_to(self, min_tokens_to_keep, token_id, mask = None):
        ...
    def gen_prune_left(self, num_tokens, mask = None):
        ...
    def gen_num_tokens(self):
        ...
    def generate_simple(self, prompt, max_new_tokens = 128):
        ...
    def apply_rep_penalty(self, logits):
        ...
    def gen_single_token(self, constraints = None, mask = None):
        ...
    class Beam:
        ...
    def begin_beam_search(self):
        ...
    def beam_search(self):
        ...
    def end_beam_search(self):
        if not self.in_beam_search: return
        self.in_beam_search = False
    def replace_last_token(self, token, seq = False):
        ...
    def sequence_ends_with(self, tokens):
        ...
'''
\end{lstlisting}

\newpage

\begin{table*}
\centering
\resizebox{\linewidth}{!}
{
\begin{tabular}{l|cccc|cccc|cccc|cccc}
    \toprule
    \multirow{3}{*}{Methods} & \multicolumn{4}{c|}{CodeGen-350M} & \multicolumn{4}{c|}{SantaCoder-1.1B} & \multicolumn{4}{c|}{CodeGen25-7B} & \multicolumn{4}{c}{StarCoder-15.5B} \\
    \cmidrule(lr){2-5} \cmidrule(lr){6-9} \cmidrule(lr){10-13} \cmidrule(lr){14-17}
    & EM & ES & ID.EM & F1 & EM & ES & ID.EM & F1 & EM & ES & ID.EM & F1 & EM & ES & ID.EM & F1 \\
    
    \midrule
    \methodname            & \textbf{22.12} & \textbf{60.41} & \textbf{29.73} & \textbf{46.09} & \textbf{30.26} & \textbf{66.90} & \textbf{39.08} & \textbf{55.43} & \textbf{36.46} & \textbf{70.76} & \textbf{44.67} & \textbf{60.40} & \textbf{46.49} & \textbf{76.80} & \textbf{55.98} & \textbf{70.32} \\

    \quad w/o cross\_df    & 19.75 & 58.95 & 27.19 & 43.52 & 27.05 & 65.12 & 35.61 & 52.23 & 32.95 & 68.97 & 40.89 & 56.97 & 42.01 & 74.40 & 51.21 & 65.89 \\
    \quad w/o intra\_df    & 16.67 & 57.28 & 23.62 & 40.11 & 23.03 & 62.87 & 31.09 & 47.89 & 27.83 & 66.42 & 35.66 & 52.25 & 43.88 & 75.39 & 53.07 & 67.62 \\
    \quad w/o dataflow    & 15.45 & 56.40 & 22.33 & 38.73 & 21.58 & 62.01 & 29.62 & 46.44 & 26.42 & 65.65 & 34.14 & 50.67 & 40.46 & 73.63 & 49.45 & 64.37 \\
    
    \bottomrule
\end{tabular}}
\caption{Ablation study for dataflow analysis on the \recceval dataset.}
\label{tab:abl-recceval}
\end{table*}

\begin{table*}
\centering
\resizebox{\linewidth}{!}
{
\begin{tabular}{l|cccc|cccc|cccc}
    \toprule
    \multirow{3}{*}{Methods} & \multicolumn{4}{c|}{2K} & \multicolumn{4}{c|}{4K} & \multicolumn{4}{c}{8K} \\
    \cmidrule(lr){2-5} \cmidrule(lr){6-9} \cmidrule(lr){10-13}
    & EM & ES & ID.EM & F1 & EM & ES & ID.EM & F1 & EM & ES & ID.EM & F1 \\

    \midrule
    Zero-Shot   & \ \ 9.49 & 61.97 & 16.44 & 47.36 & \ \ 9.76 & 62.17 & 16.70 & 47.51 & \ \ 9.61 & 62.14 & 16.59 & 47.51 \\
    CCFinder-1  & 17.11 & 66.28 & 26.02 & 55.34 & 18.12 & 66.99 & 27.17 & 56.54 & 17.79 & 67.28 & 27.05 & 56.65 \\
    RG-1        & 17.82 & 67.46 & 27.43 & 56.51 & 21.88 & 69.60 & 31.44 & 59.65 & 25.10 & 71.84 & 36.06 & 63.27 \\
    RepoCoder   & 19.10 & \textbf{68.96} & 29.83 & 58.77 & \textbf{24.24} & \textbf{71.29} & \textbf{35.01} & \textbf{62.39} & \textbf{28.29} & \textbf{73.51} & \textbf{39.44} & \textbf{65.24} \\
    
    \methodname & \textbf{21.35} & 68.78 & \textbf{30.66} & \textbf{59.08} & 20.94 & 68.65 & 30.09 & 58.68 & 20.08 & 68.42 & 29.12 & 58.32 \\
    \bottomrule
\end{tabular}
}
\caption{Performance comparison of the Code Llama-7B model (with 2K, 4K, 8K maximum input lengths) on the \crosscodeeval dataset.}
\label{tab:exp-codellama-cceval}
\end{table*}

\begin{table*}
\centering
\resizebox{\linewidth}{!}
{
\begin{tabular}{l|cccc|cccc|cccc}
    \toprule
    \multirow{3}{*}{Methods} & \multicolumn{4}{c|}{2K} & \multicolumn{4}{c|}{4K} & \multicolumn{4}{c}{8K} \\
    \cmidrule(lr){2-5} \cmidrule(lr){6-9} \cmidrule(lr){10-13}
    & EM & ES & ID.EM & F1 & EM & ES & ID.EM & F1 & EM & ES & ID.EM & F1 \\

    \midrule
    Zero-Shot   & 13.85 & 58.82 & 20.60 & 38.01 & 13.93 & 58.96 & 20.74 & 38.14 & 13.93 & 59.04 & 20.74 & 38.26 \\
    CCFinder-1  & 26.51 & 65.92 & 34.62 & 51.08 & 28.06 & 66.95 & 36.33 & 52.68 & 28.01 & 67.21 & 36.12 & 52.74 \\
    RG-1        & 30.55 & 67.90 & 37.78 & 52.66 & 36.64 & 71.15 & 44.44 & 58.22 & 41.65 & 73.95 & 49.64 & 63.05 \\
    RepoCoder   & \textbf{34.10} & \textbf{69.65} & \textbf{41.50} & 56.17 & \textbf{40.55} & \textbf{72.97} & \textbf{48.38} & \textbf{61.49} & \textbf{44.76} & \textbf{75.31} & \textbf{52.28} & \textbf{65.43} \\
    
    \methodname & 31.54 & 68.87 & 40.02 & \textbf{56.26} & 33.25 & 69.68 & 42.02 & 58.06 & 30.04 & 68.06 & 38.26 & 54.58 \\
    \bottomrule
\end{tabular}
}
\caption{Performance comparison of the Code Llama-7B model (with 2K, 4K, 8K maximum input lengths) on the \recceval dataset.}
\label{tab:exp-codellama-recceval}
\end{table*}

\begin{table*}
\centering
\resizebox{\linewidth}{!}
{
\begin{tabular}{l|cccc|cccc|cccc}
    \toprule
    \multirow{3}{*}{Methods} & \multicolumn{4}{c|}{2K} & \multicolumn{4}{c|}{4K} & \multicolumn{4}{c}{8K} \\
    \cmidrule(lr){2-5} \cmidrule(lr){6-9} \cmidrule(lr){10-13}
    & EM & ES & ID.EM & F1 & EM & ES & ID.EM & F1 & EM & ES & ID.EM & F1 \\

    \midrule
    Zero-Shot   & \ \ 8.71 & 61.95 & 16.02 & 47.51 & \ \ 8.74 & 62.07 & 16.06 & 47.63 & \ \ 8.71 & 62.08 & 16.02 & 47.58 \\
    CCFinder-1  & 22.51 & 69.15 & 31.82 & 59.77 & 26.08 & 71.29 & 35.83 & 62.80 & 27.99 & 72.59 & 38.24 & 64.46 \\
    RG-1        & 18.31 & 68.25 & 28.37 & 57.27 & 21.69 & 70.50 & 31.97 & 60.53 & 26.27 & 72.70 & 37.00 & 64.04 \\
    RepoCoder   & 20.26 & 69.84 & 30.51 & 59.31 & 24.77 & 72.69 & 36.25 & 63.57 & 29.12 & 74.56 & 40.83 & 66.81 \\
    
    \methodname & \textbf{28.78} & \textbf{72.39} & \textbf{38.72 }& \textbf{64.90} & \textbf{32.80} & \textbf{74.89} & \textbf{43.49} & \textbf{68.42} & \textbf{34.67} & \textbf{75.83} & \textbf{45.63} & \textbf{69.93} \\
    \bottomrule
\end{tabular}
}
\caption{Performance comparison of the StarCoder-15.5B model (with 2K, 4K, 8K maximum input lengths) on the \crosscodeeval dataset.}
\label{tab:exp-starcoder-cceval}
\end{table*}

\begin{table*}
\centering
\resizebox{\linewidth}{!}
{
\begin{tabular}{l|cccc|cccc|cccc}
    \toprule
    \multirow{3}{*}{Methods} & \multicolumn{4}{c|}{2K} & \multicolumn{4}{c|}{4K} & \multicolumn{4}{c}{8K} \\
    \cmidrule(lr){2-5} \cmidrule(lr){6-9} \cmidrule(lr){10-13}
    & EM & ES & ID.EM & F1 & EM & ES & ID.EM & F1 & EM & ES & ID.EM & F1 \\

    \midrule
    Zero-Shot   & 12.55 & 58.65 & 19.72 & 37.89 & 12.85 & 58.80 & 20.03 & 38.14 & 12.77 & 58.84 & 20.03 & 38.12 \\
    CCFinder-1  & 30.69 & 68.35 & 39.13 & 55.32 & 36.19 & 71.13 & 44.70 & 60.21 & 39.33 & 73.05 & 48.18 & 63.49 \\
    RG-1        & 30.78 & 68.33 & 38.23 & 53.34 & 36.88 & 71.64 & 44.78 & 59.06 & 42.67 & 74.64 & 51.11 & 64.64 \\
    RepoCoder   & 34.62 & 70.39 & 42.41 & 56.94 & 40.35 & 73.32 & 48.26 & 62.37 & 46.26 & 76.44 & 54.47 & 67.59 \\
    
    \methodname & \textbf{39.61} & \textbf{73.32} & \textbf{48.85} & \textbf{64.44} & \textbf{44.03} & \textbf{75.42} & \textbf{53.34} & \textbf{68.26} & \textbf{46.49} & \textbf{76.80} & \textbf{55.98} & \textbf{70.32} \\
    \bottomrule
\end{tabular}
}
\caption{Performance comparison of the StarCoder-15.5B model (with 2K, 4K, 8K maximum input lengths) on the \recceval dataset.}
\label{tab:exp-starcoder-recceval}
\end{table*}

\end{document}